\documentclass[%
 reprint,
superscriptaddress,
showpacs,
 amsmath,amssymb,
aps,
pra,
]{revtex4-1}

\usepackage{graphicx}
\usepackage{dcolumn}
\usepackage{bm}

\usepackage{braket}
\usepackage{color}
\usepackage[caption=false]{subfig}
\begin{document}

\preprint{APS/123-QED}

\title{Environment-Protected Solid State Based Distributed Charge Qubit}

\author{Amin Tayebi}
\email{tayebiam@msu.edu}
\affiliation{Department of Electrical and Computer Engineering, College of Engineering, Michigan State University, East Lansing, Michigan 48824, USA}
\affiliation{Department of Physics and Astronomy, Michigan State University, East Lansing, Michigan 48824, USA}
\author{Tanya Nicole Hoatson}%
\affiliation{Stanford University, Stanford, California 94305, USA}
\author{Joie Wang}%
\affiliation{University of Chicago, Chicago, Illinois 60637, USA}
\author{Vladimir Zelevinsky}%
\affiliation{Department of Physics and Astronomy, Michigan State University, East Lansing, Michigan 48824, USA}
\affiliation{National Superconducting Cyclotron Laboratory, Michigan State University, East Lansing, Michigan 48824, USA}

\date{\today}

\begin{abstract}
A novel solid state based charge qubit is presented. 
The system consists of a one-dimensional wire with a pair of qubits embedded at its center. It is shown that the system supports collective states localized in the left and right sides of the wire and therefore, as a whole, performs as a single qubit. The couplings between the ground and excited states of the two central qubits are inversely proportional making them fully asynchronized and allowing for coherent manipulation and gate operations. Initialization and measurement devices, such as leads and charge detectors, connected to the edges of the wire are modeled by a continuum of energy states. The coupling to the continuum is discussed using the effective non-Hermitian Hamiltonian. At weak continuum coupling, all internal states uniformly acquire small decay widths. This changes dramatically as the coupling strength increases: the width distribution undergoes a sharp restructuring and is no longer uniformly divided among the eigenstates. Two broad resonances localized at the ends of the wire are formed. These superradiant states (analogous to Dicke states in quantum optics), effectively protect the remaining internal states from decaying into the continuum and hence increase the lifetime of the qubit. Environmental noise is introduced by considering random Gaussian fluctuations of electronic energies. The interplay between decoherence and superradiance is studied by solving the stochastic Liouville equation. In addition to increasing the lifetime, 
the emergence of the superradiant states increases the qubit coherence.
\end{abstract}

\maketitle


\section{Introduction} \label{introduction}

Despite great theoretical advances in the field of quantum information, the best system to realize the underlying physical layer of quantum computers is still a matter of debate. The problem arises with the most fundamental element of the quantum computer, a single qubit. Ideally, error-free qubits are desired due to the fact that unknown quantum states cannot be replicated without loss of information, a consequence of the so-called no-cloning theorem. Even with advances in error correcting codes and fault-tolerant quantum computing, we still require qubits which are inherently resistive to noise and decoherence and therefore have long life and coherence times capable of outlasting gate operation timespans.

Over the past two decades, many systems have been proposed for the physical implementation of a qubit. Some examples are photonic qubits in optical quantum computers \cite{opticQC1,opticQC2}, collective spin states in nuclear magnetic resonance (NMR) systems \cite{NMRQC1,NMRQC2,NMRQC3}, electronic states in trapped ions \cite{TrapIonQC1, TrapIonQC2, TrapIonQC3}, charge, flux or phase qubits in superconducting circuits \cite {SPCQ1, SPCQ2, SPCQ3, SPCQ4}, electronic states on the surface of superfluid helium \cite{Dykman0, Dykman} and electronic charge or spin in solid state systems \cite{SSQC1, SSQC2, SSQC3}. For a comprehensive review and discussion of advantages and disadvantages of various qubit implementations see \cite{pavivcic}.

Recently, solid-state based quantum components have sparked great interest, mainly due to their scalability. In addition, the  accumulated knowledge in nano-circuit fabrication, along with existing infrastructure, can be combined to more easily realize nano-scale integrated quantum computers \cite{SSQC4}.

In any physical implementation, it is critical to consider not only the system of interest but also the constituents that the system interacts with once placed in a greater final design. In the case of solid-state based quantum computing systems, the qubits interact with devices for writing and reading information in and out, such as leads and charge detectors. These devices can be modeled as a continuum due to the large density of states they possess. Consequently, one has to deal with an \emph{open} mesoscopic quantum system where the intrinsic states are coupled to the external world through a number of channels. Similarly to a nuclear or molecular reaction, each channel is characterized by the energy of the final states and their quantum numbers. The channels have energy thresholds when the coupling opens connecting the system to the environment. Such situations require a correct unified treatment of the discrete bound states and the continuum.

A convenient framework for such problems is given by the effective non-Hermitian Hamiltonian approach introduced by Feshbach \cite{Feshbach}. The description is non-perturbative and formally exact, treating the dynamics of systems with weak, intermediate and strong couplings to the environment on equal footing. This framework is quite flexible and can be adjusted to various types of problems ranging from nuclear reactions \cite{SOKOLOV1, SOKOLOV2} to electronic transport in mesoscopic physics \cite{Tayebi1, Ziletti} and quantum optics \cite{Greenberg}. An overview of the approach and some of its results can be found in \cite{Auerbach}. To stress the large spectrum of applications using this approach we note its utilization in plasmonic antenna arrays \cite{PlasmonicArray} and biological light harvesting complexes \cite{CelardoBiology}.

A simple system of a quantum wire with a two-level atom (qubit) inserted in the middle was considered using the effective Hamiltonian in \cite{Tayebi2}. The qubit was used to regulate transport in the wire. In fact, insertion of qubits in various systems was found to be useful in different applications. For instance, in \cite{Kwapi}, a charge detector is proposed consisting of a qubit attached to a quantum wire. In this paper we consider a similar system; a quantum wire with two embedded qubits which, as a whole, behave like a single qubit.

In Sec.~\ref{secII} we introduce  a \emph{closed} one-dimensional chain of $2N$ identical sites with hoping between adjacent cells. The center of the chain is occupied by an {\sl asynchronized} pair of two-level atoms: the couplings between the ground states and excited states of the two qubits are inversely proportional. The band structure of the system consists of delocalized states extended over the chain and additional states outside the Bloch band confined to the central qubits. Depending on the coupling strengths within the qubits, the states inside the band are localized in the left arm, the right arm, or evenly distributed across the entire chain. Consequently the system acts as a \emph{distributed} charge qubit with collective right and left states.

Sec.~\ref{secIII} considers the \emph{open} system with the edge sites coupled to the continuum. A brief description of the non-Hermitian Hamiltonian approach is provided. The continuum represents ideal leads or charge detectors attached to the edges. Due to the interaction with the leads the energy states acquire decay widths influenced by the continuum coupling in a non-trivial manner. At sufficiently strong coupling, protective superradiant edge states are formed. These states steal the entire width and hence prevent the remaining states from decaying into the environment.

In Sec.~\ref{secIV} we perform a numerical study on the effect of noise and investigate the phenomenon of decoherence. The noise in the environment is modeled by a random Gaussian process. The situation is described by the stochastic Liouville equation that determines the evolution of the density matrix. It is shown that the formation of superradiant states can maximize the coherence time of the distributed qubit system.

Sec.~\ref{secV} includes the summary, concluding remarks and outlook for future work.

\section{Closed system} \label{secII}

The proposed system consists of a nano-wire with a pair of two-level atoms embedded at its center. The wire is considered to be a one-dimensional chain of $2N$ identical sites numbered as $n=-N, -(N-1),..., -1$ and $n=1,..., N-1, N$. The chain is modeled by a tight-binding Hamiltonian with coupling between adjacent neighbors:
\begin{equation} \label{H_wire}
H_{\textrm{W}}=\sum_{\substack{n=-N \\ n\neq 0}}^{N} \epsilon_0 c^{\dagger}_{n}c_{n}+\sum_{n, n'}\nu \big( c^{\dagger}_{n}c_{n'}+c^{\dagger}_{n'}c_{n} \big),
\end{equation}
where $\epsilon_0$ is the on-site energy, $\nu$ is the hopping integral, 
and $c^{\dagger}_{n}$ and $c_{n}$ are creation and annihilation operators at site $\ket{n}$, respectively. The typical value for $\nu$ is within the range of 1-100 $\mu$eV in quantum dot systems \cite{SSQC3,expermintnu,expermintnu2}. The second sum in the Hamiltonian (\ref{H_wire}) runs over the nearest neighboring cells only.

Two asynchronized qubits are symmetrically connected to the center of the wire. The left qubit with excited state $\ket{e_L}$ and energy level $\delta_{L}$ is connected to site $\ket{-1}$ with Hamiltonian
\begin{equation} \label{H_left_qubit}
H_{Q}^{L}=\delta_{L} c^{\dagger}_L c_L+\lambda \big( c^{\dagger}_L c_{-1}+ c^{\dagger}_{-1}c_L\big),
\end{equation}
where $\lambda$ is the matrix element of the qubit excitation and $c^{\dagger}_L$ and $c_L$ are creation and annihilation operators for the left qubit excited state, respectively. Similarly, the right qubit with excited state $\ket{e_R}$ and energy level $\delta_{R}$ is connected to site $\ket{1}$
\begin{equation} \label{H_right_qubit}
H_{Q}^{R}=\delta_{R} c^{\dagger}_R c_R+\frac{\kappa}{\lambda} \big( c^{\dagger}_R c_{1}+ c^{\dagger}_{1}c_R\big).
\end{equation}
Here, $c^{\dagger}_R$ and $c_R$ create and annihilate an excitation in the qubit upper level, respectively. $\kappa/\lambda$ is the coupling strength between the ground and excited states of the two-level atom. The left and right qubits are fully asynchronized, i.e. at strong coupling between the excited and ground states of the left qubit, the two states of the right qubit are weakly coupled and vice versa; $\kappa$ is the asynchronization parameter. In practice this parameter can be tuned by introducing a local electric field and adjusting the field strength \cite{Simmons,DiCarlo}. 

The total Hamiltonian of the closed system is the sum 
\begin{equation} \label{H_total_closed_sys}
H_0=H_{\textrm{W}}+H_{L}^{Q}+H_{R}^{Q}.
\end{equation}
A generic stationary wave function of the system with energy $E$ is represented as
\begin{equation} \label{stationary_state}
\ket{\psi(E)}=\sum_{\substack{n=-N \\ n\neq 0}}^{N} a_{n}(E)\ket{n} +b_{L}(E)\ket{e_L}+b_{R}(E)\ket{e_{R}}.
\end{equation}
In order to fulfill the Schr\"{o}dinger equation, the coefficients of the superposition in (\ref{stationary_state}) satisfy the linear three-term recurrence relation
\begin{equation} \label{3termRR}
(E-\epsilon_0)a_n-\nu(a_{n-1}+a_{n+1})=0, \quad \quad \quad  n \neq 0, \pm1.
\end{equation}
The discontinuities at $\pm1$ are due to the inserted qubits in the center. This naturally divides the chain into two regions, left and right. The solutions in the two sides of the chain are given by
\begin{equation} \label{3termRRSolution}
a_{n}=\left\{%
       \begin{array}{@{}r@{{}={}}l@{\quad}r@{{\,n\,}}l@{}}
        a_{n}^{L}  & A_{L}\zeta_{+}^{n}+B_{L}\zeta_{-}^{n}, & -N \leq& <-1, \\
        a_{n}^{R} & A_{R}\zeta_{+}^{n}+B_{R}\zeta_{-}^{n}, &  1    <& \leq N, \\
\end{array}\right .
\end{equation}
where $\zeta_{\pm}$ are the roots of the characteristic polynomial of the recurrence relation,
\begin{equation}
\zeta_{\pm}=\frac{1}{2\nu}\Big[E-\epsilon_0 \pm \sqrt{(E-\epsilon_{0})^2-4\nu^2} \Big],
\end{equation}
with the obvious property $\zeta_{+}\zeta_{-}=1$. The remaining equations resulting from solving the Schr\"{o}dinger equation can be used as a boundary condition to connect the two regions in (\ref{3termRRSolution}) in order to find the constants $A_L$, $A_R$, $B_L$  and $B_R$. At $n=\pm 1$ 
we have
\begin{equation} \label{BCcenterQL1}
(E-\epsilon_0)a_{-1}-\nu(a_{-2}+a_{1})=\lambda b_L,
\end{equation}
\begin{equation}  \label{BCcenterQR1}
(E-\epsilon_0)a_{1}-\nu(a_{2}+a_{-1})=\frac{\kappa}{\lambda} b_R,
\end{equation}
\begin{equation} \label{BCcenterQL2}
(E-\delta_L)b_{L}=\lambda a_{-1},
\end{equation}
\begin{equation}  \label{BCcenterQR2}
(E-\delta_R)b_{R}=\frac{\kappa}{\lambda}a_{1}.
\end{equation}

For two important cases, the solution can be analytically obtained assuming, for simplicity, that $\delta_L=\delta_R=\delta$. In the first case, when $\lambda \rightarrow 0$, the system supports states that are either fully localized in the right side or in the left side of the chain. In the second case we consider $\lambda^2 =\kappa $, where the excitation is equally distributed between the left and the right sides of the chain. The system is analogous to a qubit where the two states are extended over the entire left or right side of the wire, with $\lambda$ serving as the coupling parameter, regulating the population in each side and enabling us to perform gate operations. In accordance, we adopt a special notation throughout the paper. The stationary state  (\ref{stationary_state}) is denoted as
\begin{equation} \label{wfNotation}
\ket{\psi(E)}=\ket{L_\lambda(E)}+\ket{R_\lambda(E)},
\end{equation}
where $\ket{L_\lambda(E)}$ contains  the components of the wave function in the left side of the chain,
\begin{equation}
\ket{L_\lambda(E)}=\sum_{n=-N}^{-1} a^L_{n}(E)\ket{n} +b_{L}(E)\ket{e_L},
\end{equation}
while $\ket{R_\lambda(E)}$ contains the components in the right side,
\begin{equation}
\ket{R_\lambda(E)}=\sum_{n=1}^{N} a^R_{n}(E)\ket{n} +b_{R}(E)\ket{e_{R}}.
\end{equation}
The subscript $\lambda$ indicates that the left and right states are changed as $\lambda$ takes different values.

\subsection{Case $\lambda \rightarrow 0$}

In the limit of very weak coupling between the two states in the left qubit, $\lambda \rightarrow 0$, the eigenstates of the system fall into two categories: states that are confined in the two central qubits and Bloch waves fully localized in the right or the left side of the chain. Since the left qubit is decoupled from the chain, there exists a state with the only non-vanishing wave function component $b_L=1$ and energy $E=\delta$. The right qubit states can be found using eqs.~(\ref{BCcenterQR1}) and~(\ref{BCcenterQR2}). The two states have energies 
\begin{equation}
E= -\,\frac{\kappa}{\lambda}\,+\,\frac{1}{2}\,(\epsilon_0+\delta)\; {\rm and} \; E =\, \frac{\kappa}{\lambda}\,+\,\frac{1}{2}\,(\epsilon_0+\delta),     
\end{equation}
corresponding to eigenstates $a_{1}=-b_{R}=1/\sqrt{2}$ and $a_{1}=b_{R}=1/\sqrt{2}$, respectively.

The states of the second type are distributed over the wire, localized either in the left side between the sites $\ket{-N}$ and $\ket{-1}$, or in the right side between $\ket{1}$ and $\ket{N}$. According to (\ref{BCcenterQR2}), since $\kappa/\lambda_c \rightarrow \infty$, we have $a_1^R=0$. This breaks the symmetrical structure of the wire, creating a longer chain on the left and leaving the right chain shorter, which is the key point in realizing a qubit structure using the proposed system. For the states in the left side (only $a_n^{L} \neq 0$), applying the boundary conditions $a_1^R=0$ and $a_{-N-1}^L=0$ to (\ref{3termRRSolution}) provides an equation for energies, $\zeta_{-}^{2N+2}=1$. Thus the energies can be parameterized by a positive even number, the quantized quasi-momentum $k$,
\begin{equation} \label{LeftTrappedEn}
E_k^L=\epsilon_0 + 2\nu \cos \varphi_k^L, \quad \varphi_k^L=\frac{\pi k}{2N+2},\quad  k~\textrm{even},
\end{equation}
and the corresponding amplitudes (\ref{3termRRSolution}) are of the Bloch-wave type,
\begin{equation}
a_n^{L}(k)=i^k \sqrt{\frac{2}{N+1}}\sin(n\varphi_k^L).
\end{equation}
In what follows, according to (\ref{wfNotation}), these states are denoted as $\ket{L_0(E)}$.

Similarly, for the states in the right side of the wire (only $a_n^{R} \neq 0$) with boundary conditions $a_1^R=0$ and $a_{N+1}^R=0$ we have $\zeta_{+}^{2N}=1$. Thus
\begin{equation} \label{RightTrappedEn}
E_{k}^{R}=\epsilon_0 + 2\nu \cos \varphi_k^R,\quad \varphi_k^R=\frac{\pi k}{2N},\quad  k~\textrm{even},
\end{equation}
with eigenfunctions
\begin{equation}
a_n^{R}(k)=i^k \sqrt{\frac{2}{N}}\sin(n\varphi_k^R).
\end{equation}
These states are denoted as $\ket{R_0(E)}$.

As $k$ varies, the energy states (\ref{LeftTrappedEn}) and (\ref{RightTrappedEn}) consecutively alternate and come in pairs. In each pair, the higher energy state is extended in the left side and the lower level in the right side, being associated with wave functions $\ket{L_0}$ and $\ket{R_0}$, respectively. Clearly, in the extreme limit of $\lambda \rightarrow \infty$ we have the reverse situation. Since $a_{-1}^L=0$, the right chain becomes longer and the upper energy state in each pair is localized in the right side of the chain and therefore can be denoted as $\ket{R_{\infty}}$. Consequently, the lower energy state is localized on the left which is indicated by $\ket{L_{\infty}}$.

\subsection{Case $\lambda^2=\kappa$}

\begin{figure}[h!]
\centering
\includegraphics[scale=0.4]{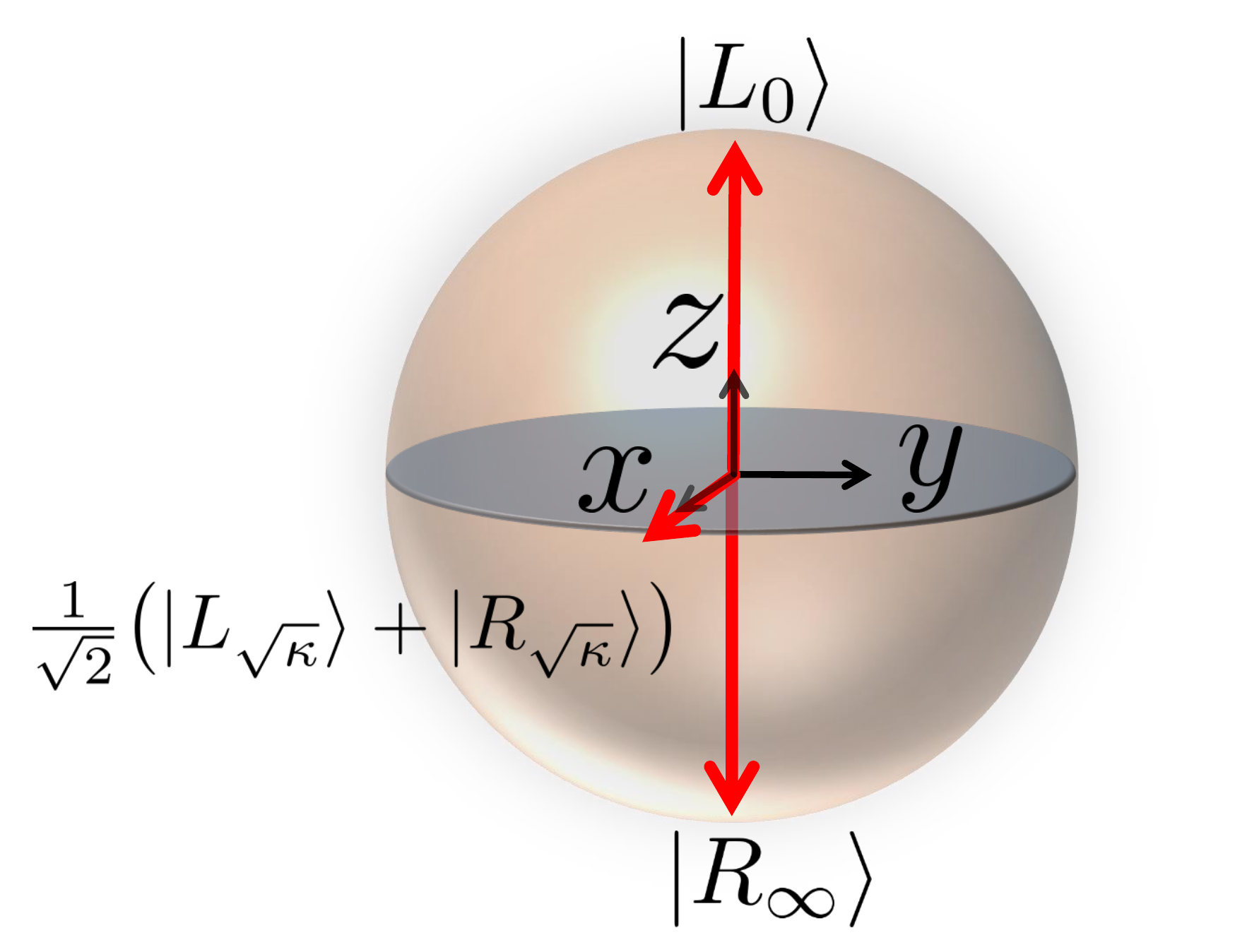}
\caption{Graphical presentation of qubit rotation on the Bloch sphere. Gate operations can be performed by varying the value of $\lambda$. The initial state at $\lambda=0$ is localized in the left side, $\ket{L_0}$. At $\lambda=\sqrt{\kappa}$ the state is 90$^\circ$ rotated becoming an equal superposition of the left and right states. When $\lambda \rightarrow \infty$, the state becomes localized on the right.}
\label{BlochSphere}
\end{figure}

For finite values of $\lambda$ the two sides of the chain are not decoupled and states are extended over the entire chain. The parameter $\lambda$ therefore acts as a knob to control the population in the left and right regions. We now consider a special case of $\lambda^2=\kappa$ where the population is equally divided between the two sides of the wire. Due to symmetry, there exist two types of eigenfunctions: fully symmetric states with $a_{n}^{L}=a_{n}^{R}$ and anti-symmetric states with $a_{n}^{L}=-a_{n}^{R}$. Using eqs.~(\ref{BCcenterQL1}) and (\ref{BCcenterQL2}) or eqs.~(\ref{BCcenterQR1}) and (\ref{BCcenterQR2}) we obtain an equation for the energies,
\begin{equation} \label{energyeqncase2}
\frac{\sin\big[(N+1)\phi\big]}{\sin(N\phi)}= \pm 1+\frac{\kappa}{\nu(E-\delta)},
\end{equation}
where $\sin \phi=(1/2\nu)\sqrt{(E-\epsilon_0)^2-4\nu^2}$, with the positive (negative) sign corresponding to symmetric (anti-symmetric) states. Analogous to an equal superposition of the two states in a qubit, we denote the symmetric and anti-symmetric states as $(1/\sqrt{2})(\ket{L_{\sqrt{\kappa}}}+\ket{R_{\sqrt{\kappa}}})$ and $(1/\sqrt{2})(\ket{L_{\sqrt{\kappa}}}-\ket{R_{\sqrt{\kappa}}})$, respectively. The evolution with varying $\lambda$ from zero to infinity and controlling the state of the qubit is graphically shown in Fig.~\ref{BlochSphere} using the Bloch sphere. Assuming that at $\lambda=0$ the state is localized on the left, $\ket{L_0}$, a 90$^\circ$ rotation ($\pi/2$ pulse) is performed by adiabatically moving $\lambda$ to  $\lambda=\sqrt{\kappa}$. Further increasing $\lambda$ to extreme values localizes the particle in the right side of the chain, $\ket{R_{\infty}}$.

\begin{figure}[h!]
\centering
\captionsetup[subfigure]{margin={0.92cm,0cm}}
\subfloat[]{
  \includegraphics[height=7.3cm,width=8cm]{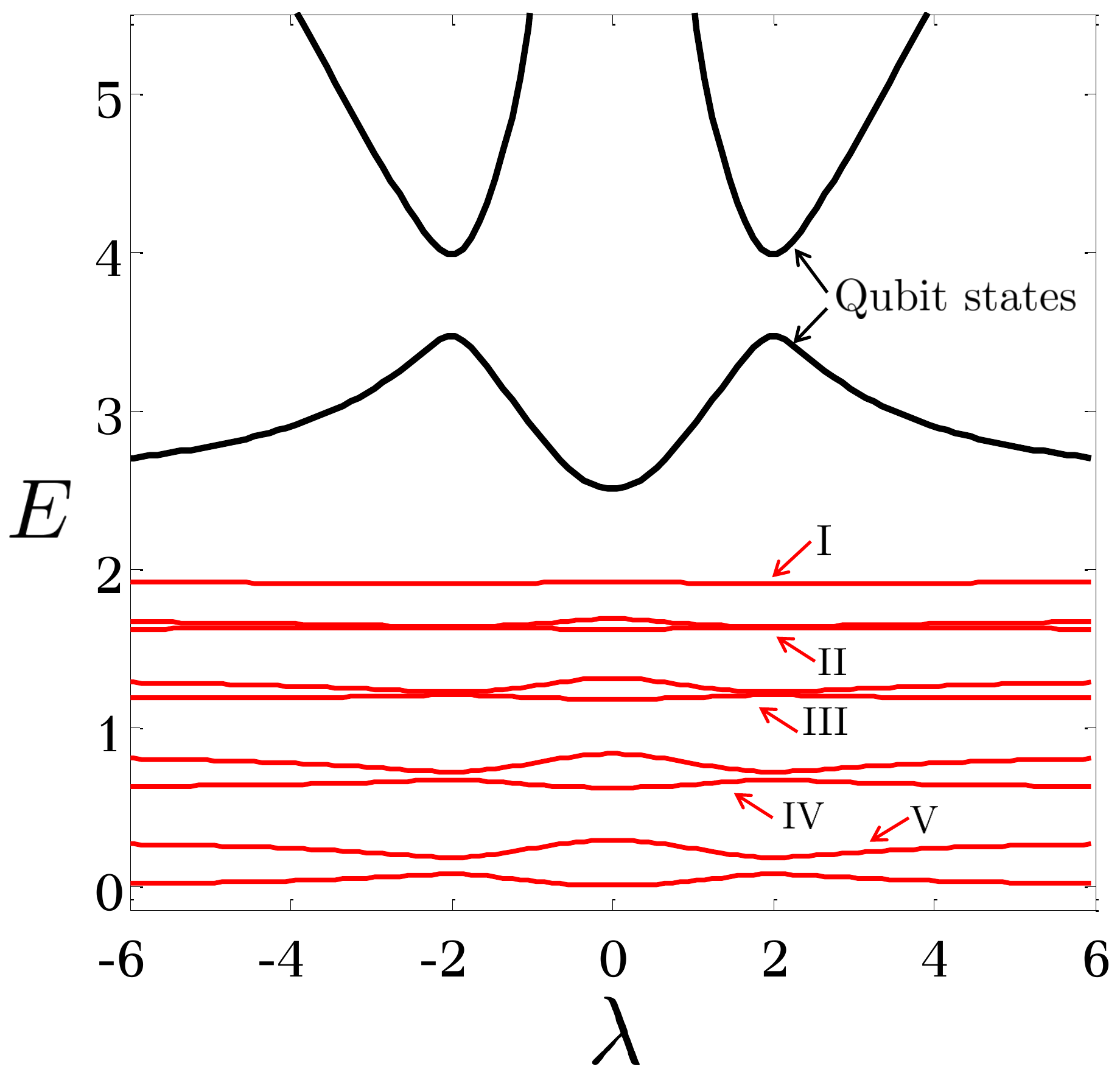}%
}

\captionsetup[subfigure]{margin={1.2cm,0cm}}
\hspace{-4mm}
\subfloat[]{%
  \includegraphics[height=3.8cm,width=8.3cm]{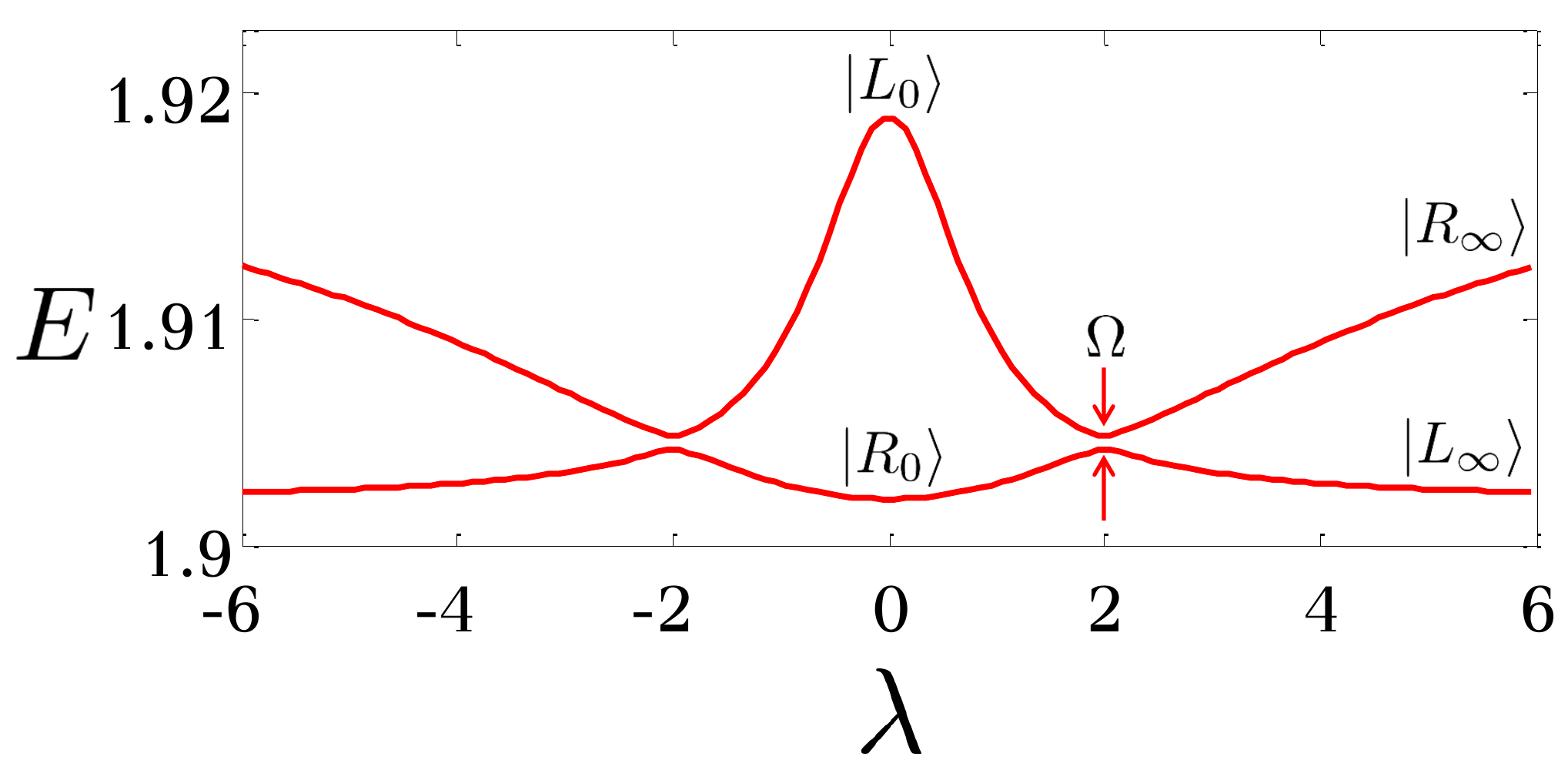}%
}

\captionsetup[subfigure]{justification=justified,singlelinecheck=false}
\caption{(a) Upper half of the band structure as a function of $\lambda$ for a system with $N=10$, $\epsilon_0=0$, $t=1$, $\delta=2.5$ and $\kappa=4$. The black curves correspond to states localized in the central qubits and the red curves are pairs of Bloch waves. (b) A closer view of pair I as a function of $\lambda$.} \label{Bandstructure}
\end{figure}

The complete band structure as a function of $\lambda$ is shown in Fig.~\ref{Bandstructure}(a). The parameters of the system are $N=10$, $\epsilon_0=0$, $\nu=1$, $\delta=2.5$ and $\kappa=4$ (for the remainder of the paper, $\epsilon_0$ is set to zero and the scale is fixed by setting $\nu=1$). The black curves correspond to states localized in the central qubits and the red curves are pairs of extended states over the wire. Fig.~\ref{Bandstructure}(b) shows pair I in a smaller energy scale. It is clear that any rotation on the Bloch sphere can be performed by varying $\lambda$. The arrow indicates the avoided crossing point, when $\lambda=\sqrt{\kappa}$. The difference between two consecutive energy solutions of eq.~(\ref{energyeqncase2}) is the Rabi frequency, $\Omega$, at which the population oscillates between the left and right sides of the chain.

\begin{figure}[h!]
\centering
\captionsetup[subfigure]{}
\subfloat[]{
  \includegraphics[scale=0.35]{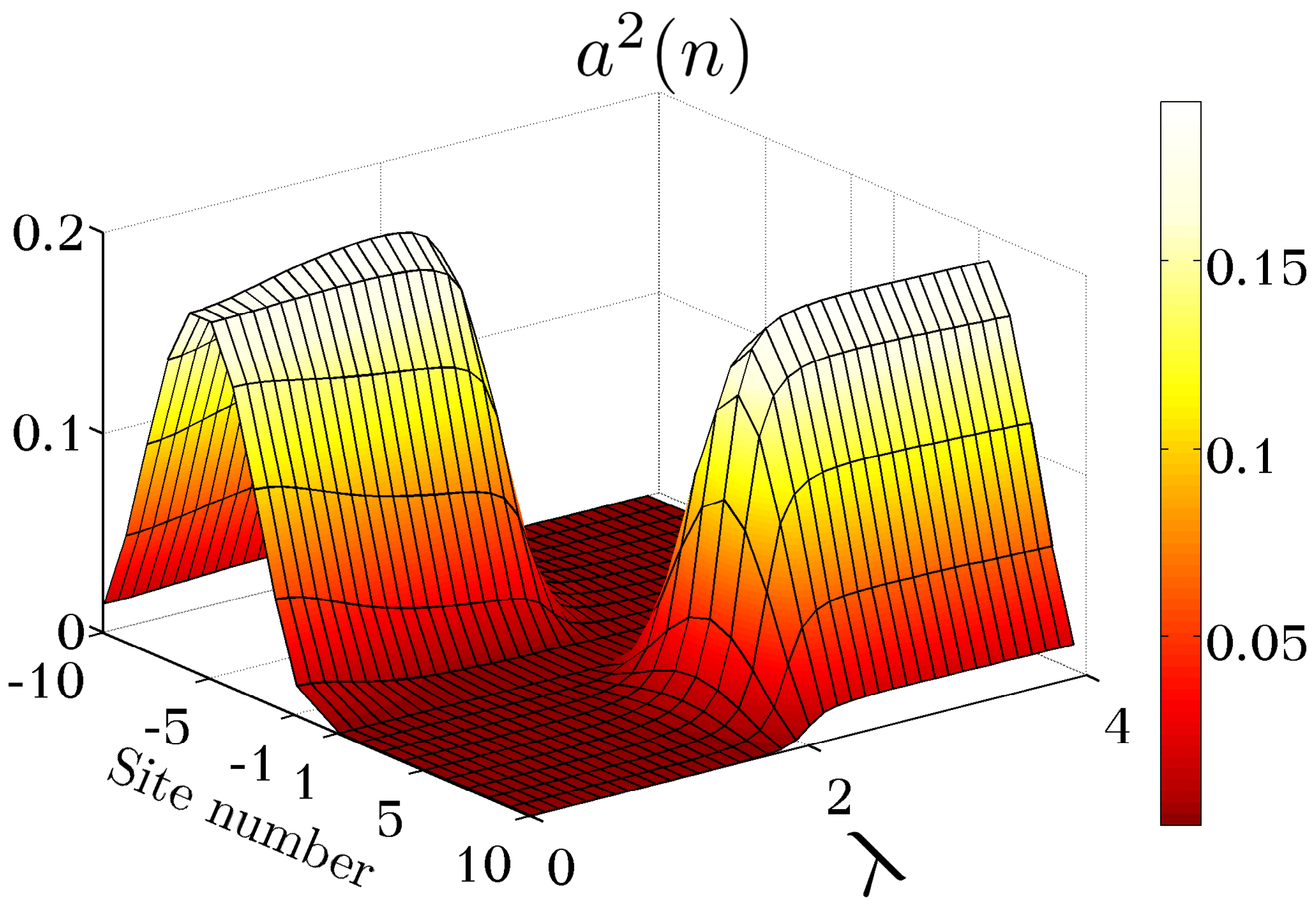}%
}
\captionsetup[subfigure]{}
\subfloat[]{%
  \includegraphics[scale=0.35]{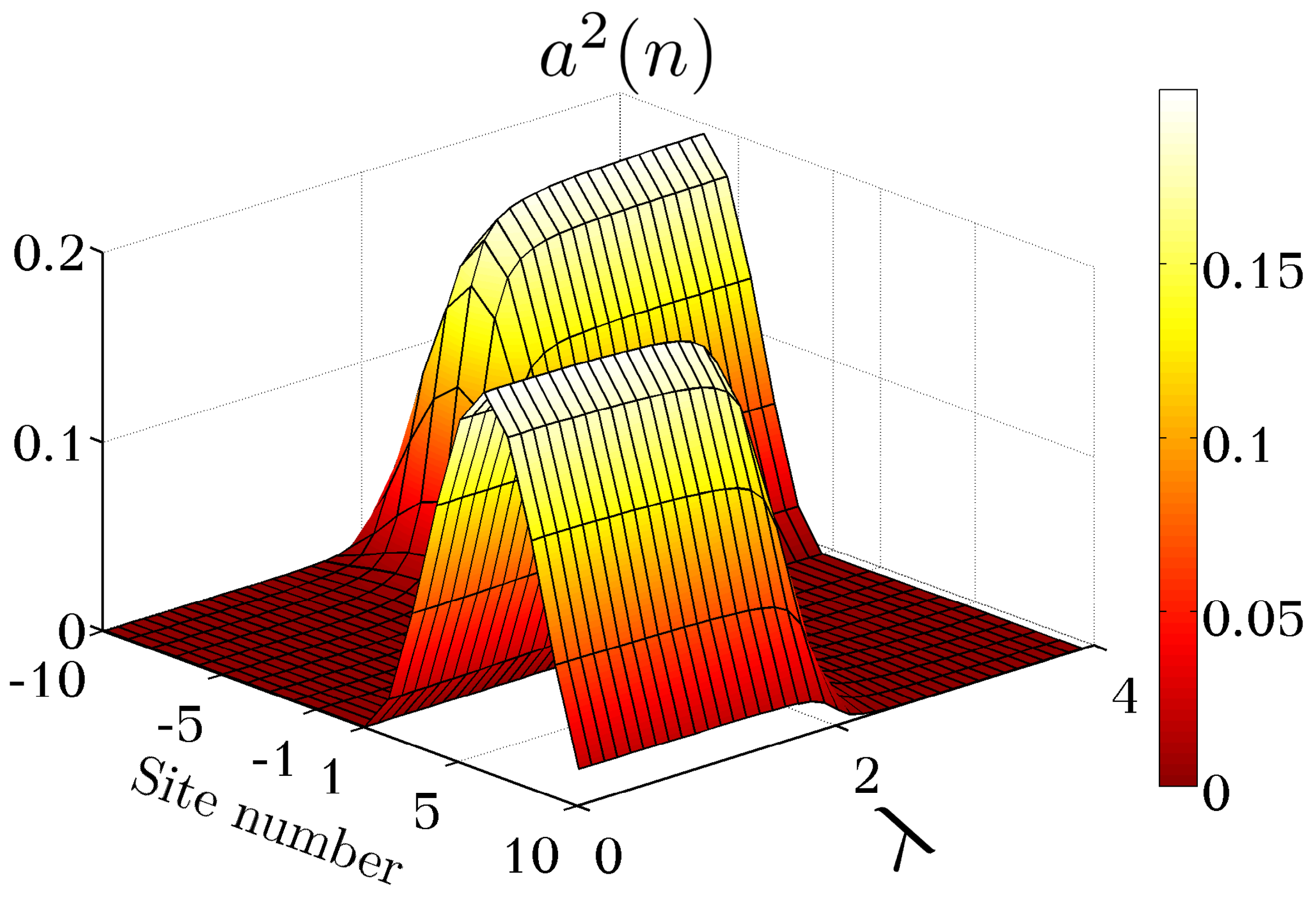}%
}
\captionsetup[subfigure]{justification=justified,singlelinecheck=false}
\caption{Evolution of the squared components of the (a) upper state and (b) lower state wave functions in pair one indicated in Fig.~\ref{Bandstructure}. Only the components in the chain are considered in the figure: amplitudes $a(n)$ in the wave function (\ref{stationary_state}).} \label{UpperandLowerP1}
\end{figure}

The squared components of the wave function (\ref{stationary_state}) of the upper and lower states in pair I as a function of $\lambda$ are shown in Fig.~\ref{UpperandLowerP1}. The figure only considers the components in the chain and not those of the excited states of the central qubits ($a_n$'s in the wave function (\ref{stationary_state})). At $\lambda=0$ the upper and lower states are fully localized in the left and right sides, respectively. At $\lambda=\sqrt{\kappa}=2$, both upper and lower states are in equal superpositions. As $\lambda$ increases, the upper state quickly becomes localized in the right side. Similarly, the lower state develops into a localized state in the left.

\begin{figure}[h!]
\centering
\includegraphics[scale=0.35]{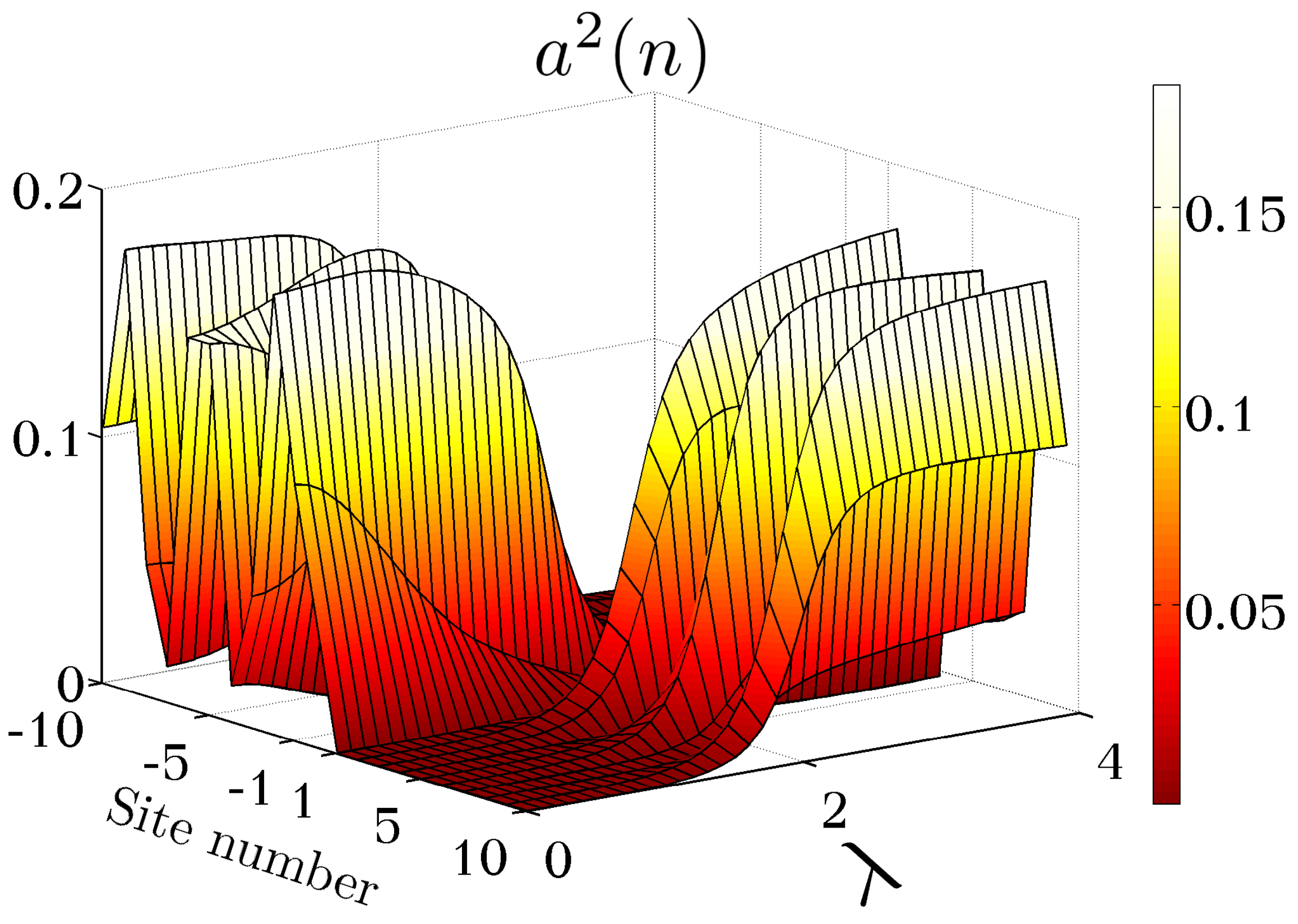}
\caption{Evolution of the squared components of the upper state wave function in pair three indicated in Fig.~\ref{Bandstructure}. Only the components inside the chain are considered in this figure.}
\label{UpperStateP3}
\end{figure}

All pairs within the Bloch band in Fig.~\ref{Bandstructure} possess a qubit structure and therefore can be considered for implementing a qubit. The wave function profile, however, varies for each pair. Fig.~\ref{UpperStateP3} shows the upper state wave function in pair III as $\lambda$ takes different values. The behavior is qualitatively the same as in pair I, when $\lambda$ increases, the state evolves from being fully localized in the left to being fully localized in the right.

\section{Open system} \label{secIII}

In this section we consider the open system with the two edge sites of the chain coupled to the external world. This represents the interaction of the system with devices such as charge detectors and leads connected to the two ends of the wire. We first briefly provide an overview of the non-Hermitian effective Hamiltonian approach. Next we apply this technique to the distributed qubit system.

\subsection{The non-Hermitian effective Hamiltonian}

Consider a quantum system with Hamiltonian $H_0$ and discrete intrinsic levels $\ket{i}$ interacting with the surrounding. 
The environment can be characterized by a continuum of channels, $\ket{c;E}$, where $E$ is the energy. The intrinsic states $|i\rangle$ are coupled to the channels with amplitudes $A_{i}^{c}(E)$. The transition amplitudes are in general energy dependent; the channel $c$ is open only if the running energy is greater than the channel energy threshold and closed otherwise. A convenient mathematical formalism describing such a system is the non-Hermitian effective Hamiltonian 
\begin{equation} \label{effectiveHamiltonian}
\mathcal{H}=H_{h}-\frac{i}{2}\,W.
\end{equation}
The effective Hamiltonian $\mathcal{H}$ living in the subspace of the closed system contains a Hermitian part, $H_{h}$, and an anti-Hermitian part, $W$. The Hermitian part, which renormalizes the energies of the closed system, is given by
\begin{equation}
H_{h}=H_{0}+\Delta(E).
\end{equation}
The matrix elements of $\Delta(E)$ between two internal states $\ket{i}$ and $\ket{j}$ are given by the Cauchy principal value integral
\begin{equation} \label{realpartHeff}
\Delta_{i,j}(E) = \sum_{c}\mathcal{P}.\mathcal{V}.\int{dE'\frac{A_{i}^{c}(E'){A_{j}^{c}}^{*}(E')}{E-E'}},
\end{equation}
while the matrix element of the anti-Hermitian part is
\begin{equation} \label{imagpartofHeff}
W_{ij}(E)= 2\pi \sum_{c_{open}} A_{i}^{c}(E){A_{j}^{c}}^{*}(E).
\end{equation}
The sum in the real part (\ref{realpartHeff}), runs over all channels, open and closed, and therefore it takes into account virtual transitions to the environment. The sum in the imaginary part (\ref{imagpartofHeff}), however, only includes contributions from real transitions to the continuum channels and hence it only runs over the channels 
open at a given energy.

In many cases, the energy interval of interest is relatively small and the transition amplitudes, $A_{i}^c(E)$, can be considered to be smooth functions of energy. Consequently, the energy dependence of the amplitudes can be neglected. Then the principal value integral in (\ref{realpartHeff}) vanishes and the effective Hamiltonian reduces to
\begin{equation} \label{ReducedHeff}
\mathcal{H}=H_{0}-\frac{i}{2}\,W.
\end{equation}
In the following subsection, the effective Hamiltonian (\ref{ReducedHeff}) is used as the starting point  for investigating the open system.

\subsection{Superradiance and emergence of protecting edge states}

Now we open our system, coupling the left-most and right-most sites ($\ket{-N}$ and $\ket{N}$) to ideal leads by the amplitudes $A_{-N}^{\ c_L}$ and $A_{N}^{c_R}$, respectively. We further assume that the energy dependence of the amplitudes can be ignored and the couplings are symmetric, $A_{-N}^{L}=A_{N}^{R}=\sqrt{\gamma}$. Here $\gamma$ is the parameter representing the interaction strength with the decay channels. The effective Hamiltonian (\ref{ReducedHeff}), with $H_0$ being the Hamiltonian of the closed system (\ref{H_total_closed_sys}), thus fully describes the situation. Since the  chain is only coupled through the edge sites, the operator $W$ takes on a simple form: according to (\ref{imagpartofHeff}) the only non-zero matrix elements of this operator are
\begin{equation}
W_{-N,-N}=W_{N,N}=\gamma.
\end{equation}

The behavior of the system is strongly influenced by 
the dimensionless parameter $\gamma/D$, where $D$ is the mean energy level spacing of the closed system. A parametric study for a chain without qubits and a chain with a single qubit was performed in \cite{Celardo} and \cite{Tayebi2}, respectively. In both cases, the typical picture was found to be as follows. At weak coupling to the environment, all intrinsic states acquire a small decay width. The width distribution among the states is almost uniform with the maximum at the center of the Bloch band. When $\gamma$ grows, the distribution abruptly changes at $\gamma \simeq D$. Beyond this point, further increasing the coupling results in the segregation of states into long-lived narrow and short-lived broad resonances. In analogy to Dicke superradiance in quantum optics \cite{DickeQO}, we term these emergent giant resonances as superradiant states [22,23].

Adopting a similar approach to our system we consider 
a chain with 20 intrinsic cells ($N=10$), a pair of asynchronized qubits attached at the center, and 
the two end sites coupled to the continuum. The results of this study are presented with a series of figures (in all figures the scale is fixed by setting $\nu=1$). 

\begin{figure}[h]
\centering
\captionsetup[subfigure]{margin={1.2cm,0cm}}
\hspace*{-0.3cm}
\subfloat[]{
  \includegraphics[scale=0.38]{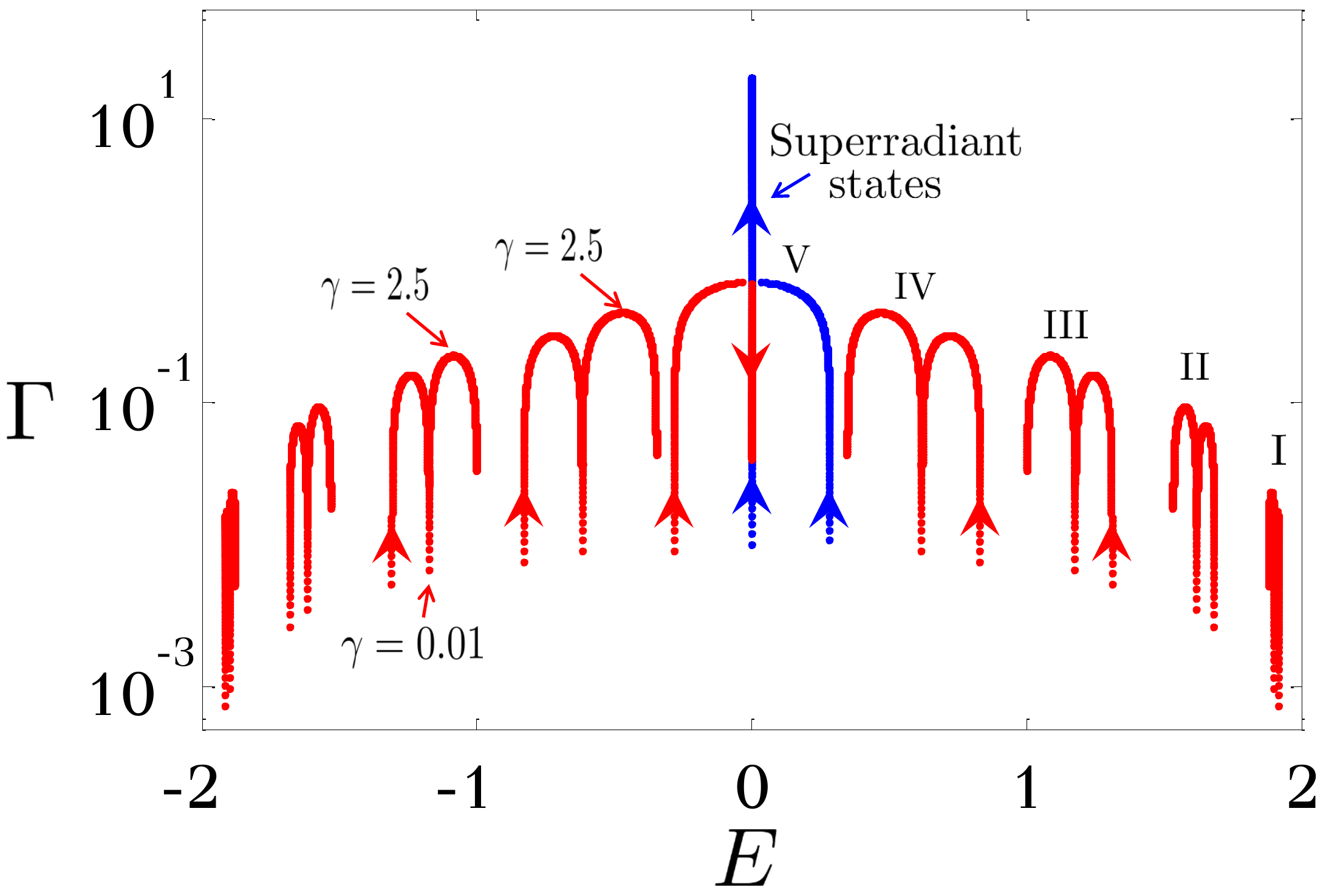}%
}
\captionsetup[subfigure]{margin={0.7cm,0cm}}
\hspace*{+0.3cm}
\subfloat[]{%
  \includegraphics[width=7.2cm,height=3.4cm]{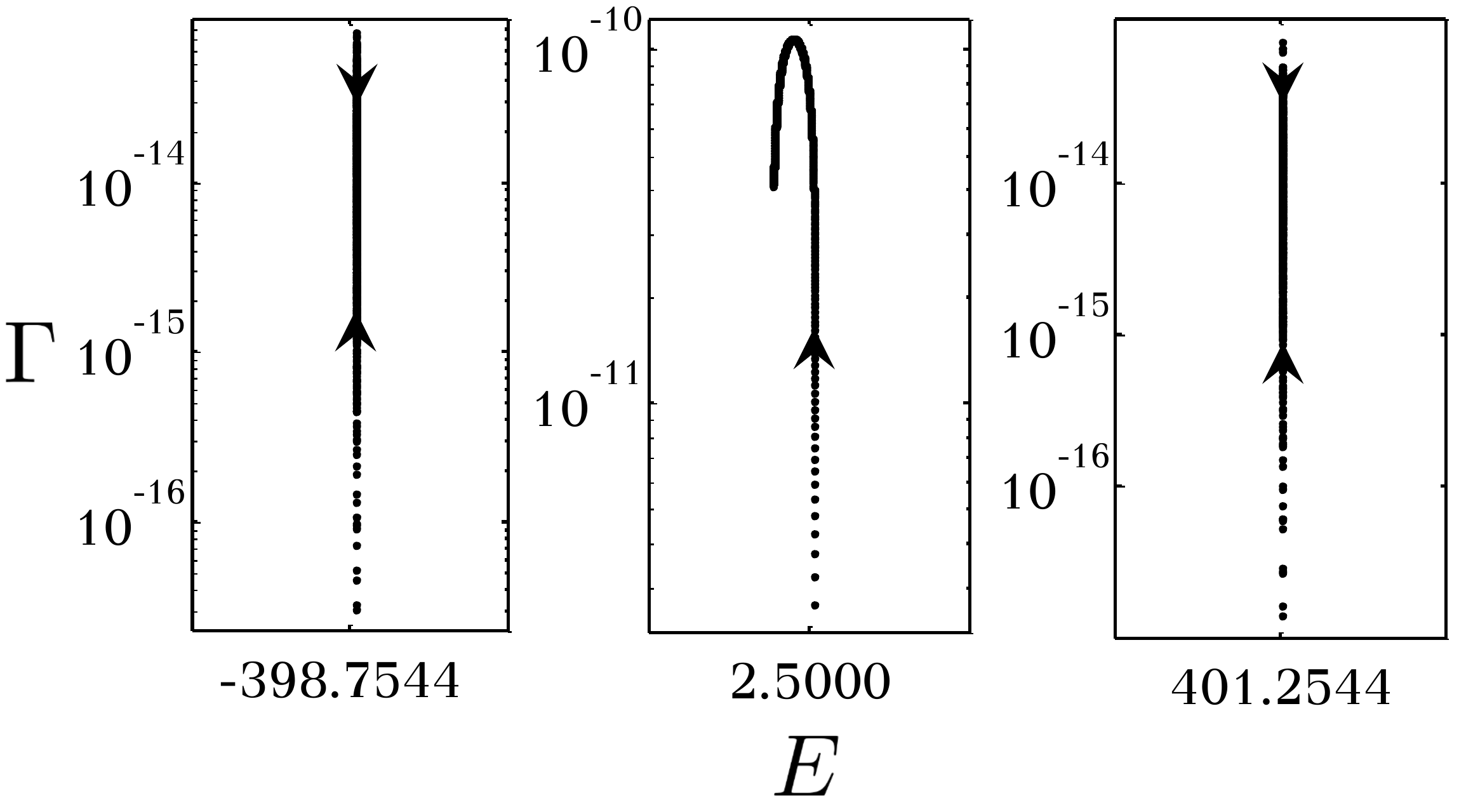}%
}

\captionsetup[subfigure]{justification=justified,singlelinecheck=false}
\caption{Evolution of complex eigenenergies as a function of coupling to the continuum for (a) states within the Bloch band and (b) states outside the band. The arrows indicate the direction in which $\gamma$ evolves from 0.05 to 20 with 0.01 increments. The parameter values are $\delta=2.5$, $\lambda=0.01$ and $\kappa=4$.} \label{ComplexPlaneevlLp01}
\end{figure}

In the first step we diagonalize the effective Hamiltonian,
\begin{equation} \label{effHamiltonSchrodinger}
\mathcal{H}\ket{q}=\mathcal{E}_q\ket{q},
\end{equation}
The trajectories of the energies $\mathcal{E}_q$ in the complex plane as a function of $\gamma$ are shown in Fig.~\ref{ComplexPlaneevlLp01}. The $x$ and $y$ axes represent the real and imaginary parts of the energies, respectively ($\mathcal{E}_q=E_{q}-(i/2)\Gamma_q$). The parameters of the system are $\delta=2.5$, $\lambda=0.01$, $\kappa=4$. States within the band are shown in the top panel. Because $\lambda$ is small, the states are fully localized in both sides of the chain. The five pairs indicated in the figure correspond to the pairs of the closed system shown in Fig.~\ref{Bandstructure}. The emergence of superradiant states is clear in the figure. At small values of $\gamma$ all states are narrow resonances. The decay widths grow as $\gamma$ increases. At the critical value of $\gamma\simeq 2.5$ the sharp superradiant transition occurs. Beyond this point, the superradiant states (pair V in the figure) become broad resonances, essentially protecting the remaining states from decaying into the continuum. The lower panel in the figure shows the states outside of the band. These states are localized in the central qubits and  have extremely small decay widths. For these states, the shift in the real energy is negligible and therefore after the superradiance transition they trace back the same trajectories.

It was noticed in \cite{vonBrentano} that the real part of the effective Hamiltonian repels the levels while attracting the widths. On the contrary, the anti-Hermitian part attracts the real energies but repels the widths. This phenomenon can be seen in Fig.~\ref{ComplexPlaneevlLp01}, where on the road to superradiance the energy levels are attracted to the center of the band.

\begin{figure}[h]
\centering
\captionsetup[subfigure]{margin={1.2cm,0cm}}
\hspace*{-0.3cm}
\subfloat[]{
  \includegraphics[scale=0.38]{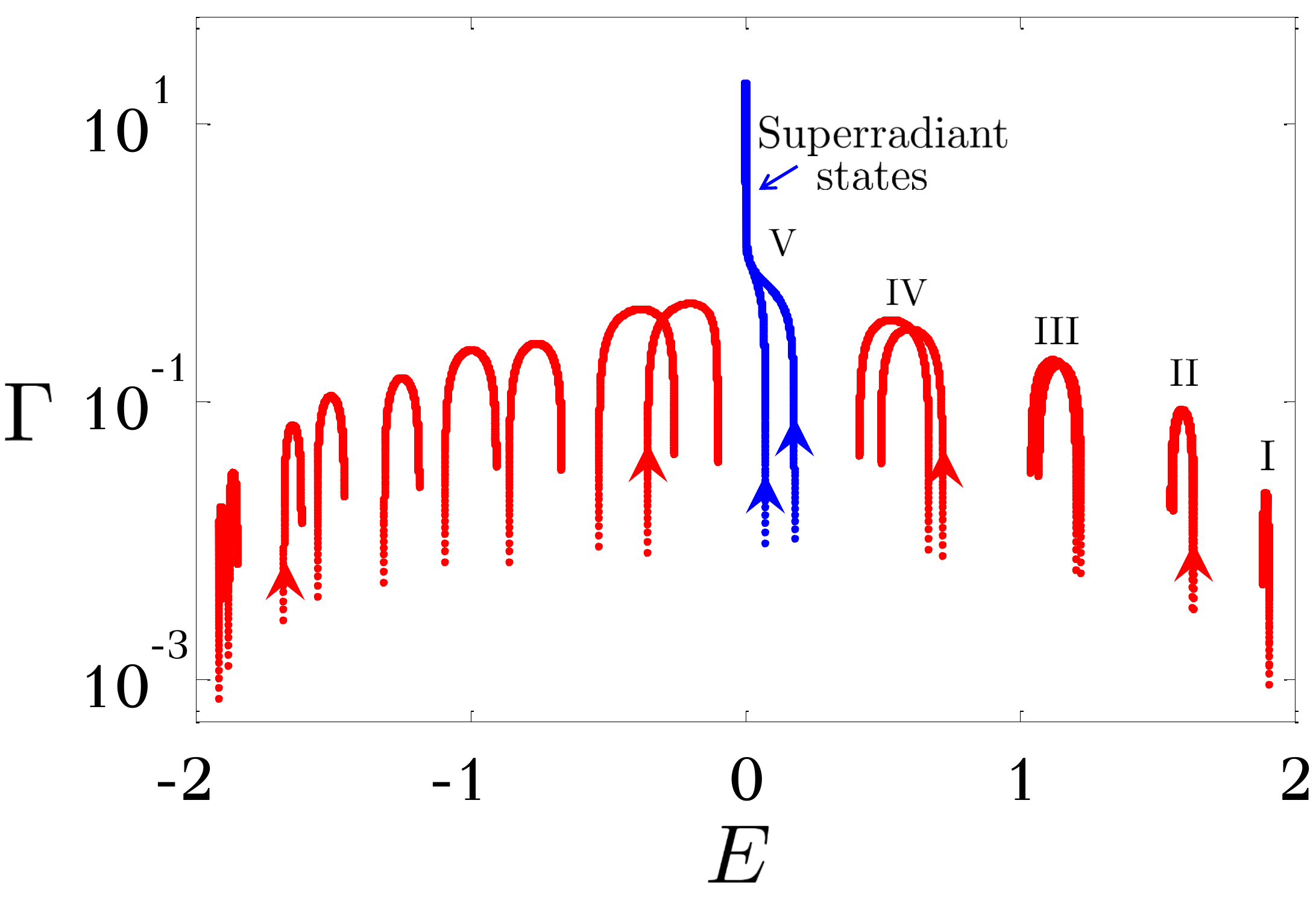}%
}
\captionsetup[subfigure]{margin={0.7cm,0cm}}
\hspace*{+0.3cm}
\subfloat[]{%
  \includegraphics[width=7.2cm,height=3.4cm]{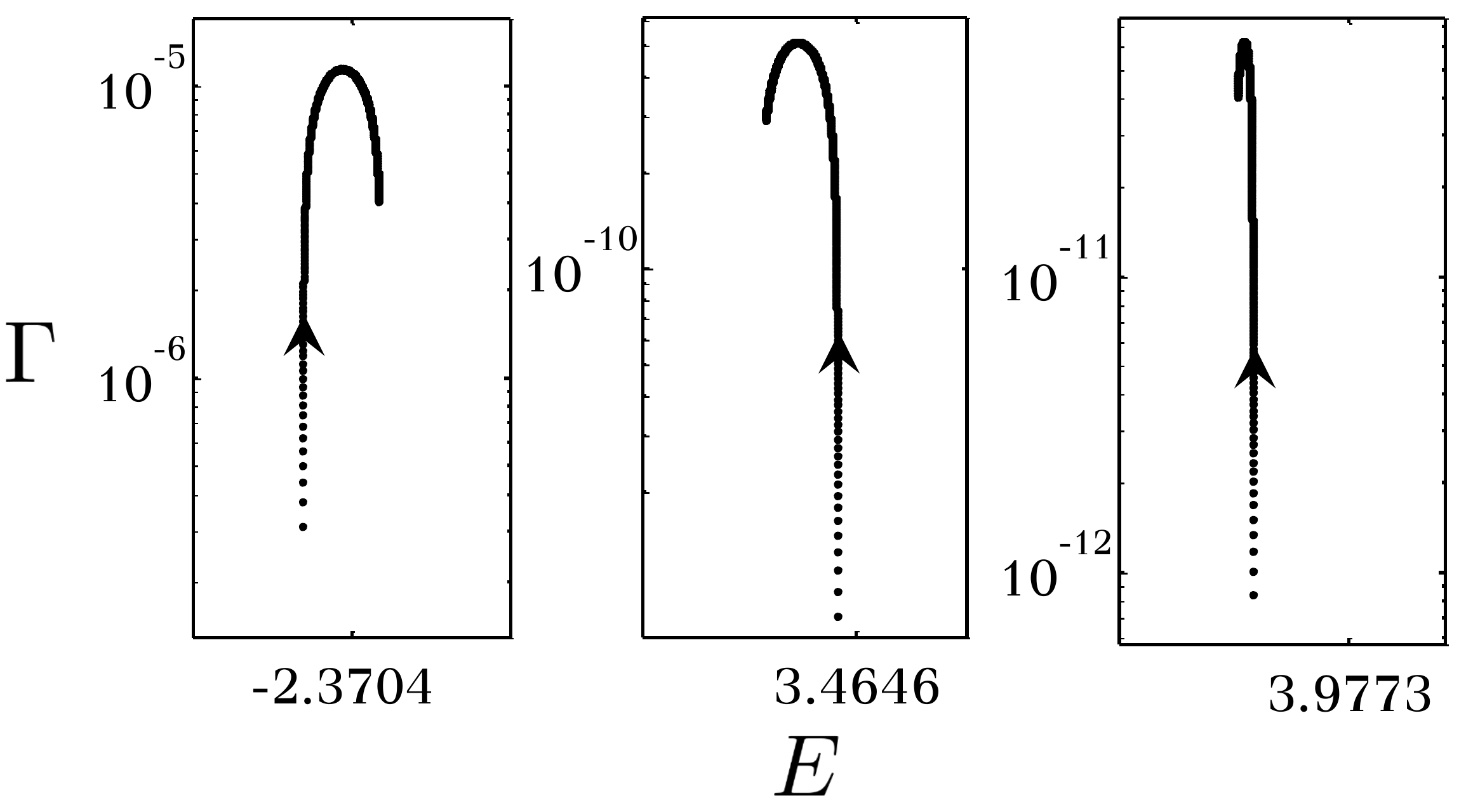}%
}
\captionsetup[subfigure]{justification=justified,singlelinecheck=false}
\caption{Evolution of the complex eigenenergies as a function of coupling to the continuum for (a) states within the Bloch band and (b) states outside the band. $\gamma$ varies from 0.05 to 20 with 0.01 increments. The parameter values are $\delta=2.5$, $\lambda=2$ and $\kappa=4$.} \label{ComplexPlaneevlL2}
\end{figure}

The evolution of complex energies of the same system when $\lambda=2$ is shown in Fig.~\ref{ComplexPlaneevlL2}. All pairs are now evenly distributed over the entire chain. The picture is qualitatively the same as in the previous case.  The two superradiant states are again placed in the center of the band. At large values of coupling to the continuum, the superradiant states acquire the entire width, leaving the remaining states to be long-lived.

In order to better envision the protecting role of the superradiant states we compare the lifetime of a particle inside the chain for different initial conditions. In the first case, the particle is initialized in the upper energy state of pair I shown in Fig.~\ref{ComplexPlaneevlL2}. In the second case, the initial state is that of the upper energy of pair V (superradiant pair) shown in the same figure. We calculate the survival probability $P(t)$ according to
\begin{equation}
P(t)= \sum_{m} \big|\langle m \ket{\psi(t)} \big|^2,
\end{equation}
where $\ket{m}$'s are the intrinsic states of the closed system ($\{\ket{n}\}$, $\ket{e_L}$ and $\ket{e_R}$ in the wave function (\ref{stationary_state})) and $\psi(t)$ is the result of quantum evolution,
\begin{equation}
\psi(t)=e^{-i \mathcal{H}_{\textrm{eff}}t}\psi_{0},
\end{equation}
where $\psi_{0}$ is the initial state. The initial state can be expanded in the biorthogonal space of the eigenstates of the effective Hamiltonian (\ref{effHamiltonSchrodinger}),  $\psi_{0}=\sum_{q} c_q \ket{q}$. Consequently we have
\begin{equation}
P(t)=\Big| \sum_{m,q} c_q e^{-i \mathcal{E}_q t}\langle m\ket{q} \Big|^2.
\end{equation}

\begin{figure}[h]
\centering
\captionsetup[subfigure]{margin={0.6cm,0cm}}
\hspace*{-1cm}
\subfloat[]{
  \includegraphics[scale=0.25]{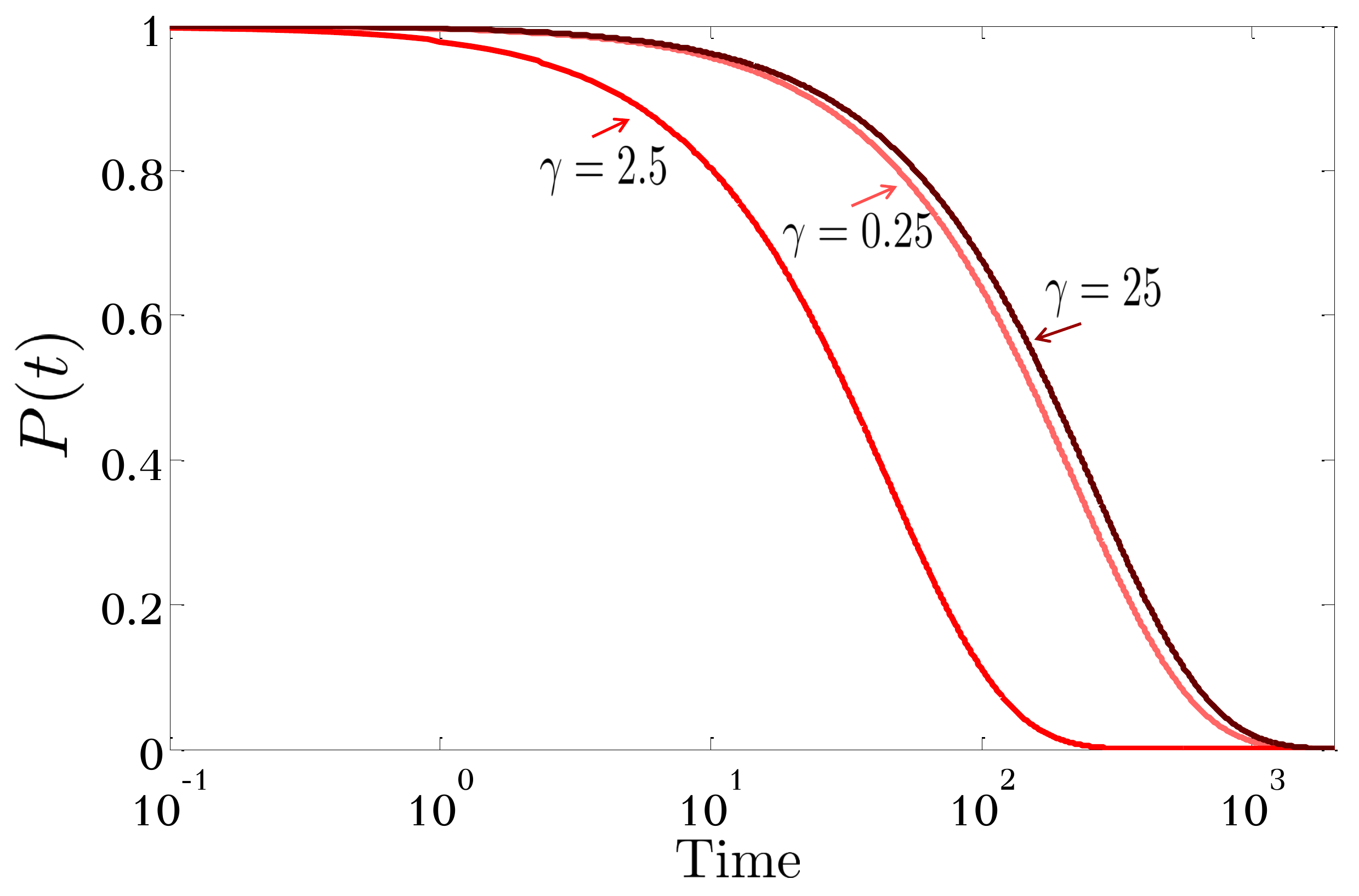}%
}
\captionsetup[subfigure]{margin={0.6cm,0cm}}
\hspace*{-1cm}
\subfloat[]{%
  \includegraphics[scale=0.25]{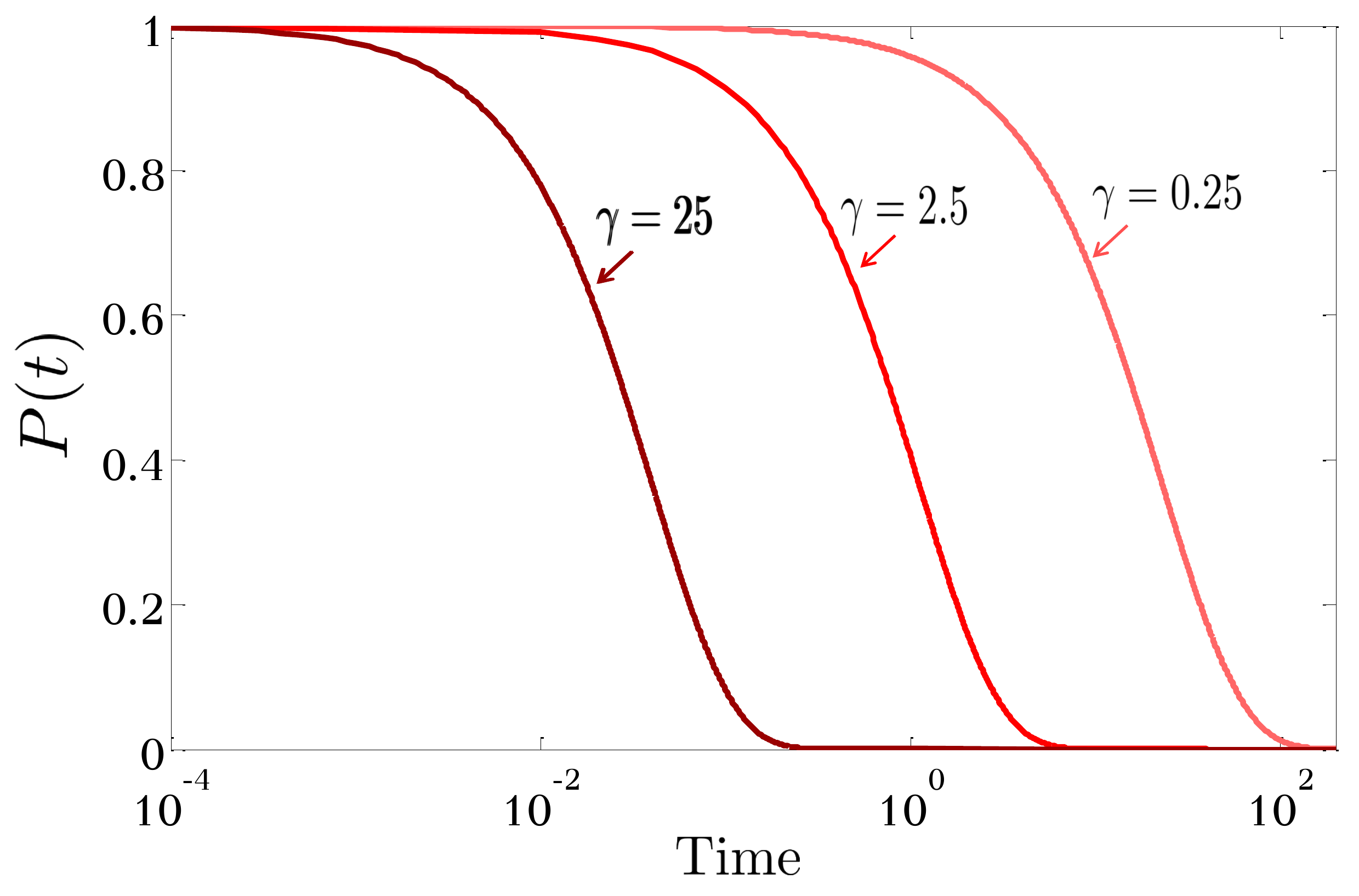}%
}
\captionsetup[subfigure]{justification=justified,singlelinecheck=false}
\caption{Lifetime for different values of coupling to the continuum $\gamma$ for a particle initialized in (a) the upper state of pair I  and (b) the upper state of pair V (superradiant pair) in Fig.~\ref{ComplexPlaneevlL2}. The parameter values are $\delta=2.5$, $\lambda=2$ and $\kappa=4$.} \label{Lifetime}
\end{figure}

The results for different values of the coupling constant $\gamma$ are shown in Fig.~\ref{Lifetime}. In panel~(a) the particle was initialized in the upper state of the first energy pair in the band structure. Compared to the weak coupling case ($\gamma=0.25$), at the superradiance transition ($\gamma=2.5$) the lifetime is by an order of magnitude smaller, which makes the situation convenient for measurement and fast readout of the qubit system. At strong coupling and beyond the superradiance transition, when $\gamma=25$, the lifetime increases and allows for storing information or performing operations on the qubit. Contrary to this, it is shown in panel~(b) that the lifetime of a particle initialized in the superradiant state monotonically decreases as the coupling strength is increased. Consequently, other pairs in the band structure become protected from decaying into the leads.

\begin{figure}[h]
\centering
\includegraphics[scale=0.4]{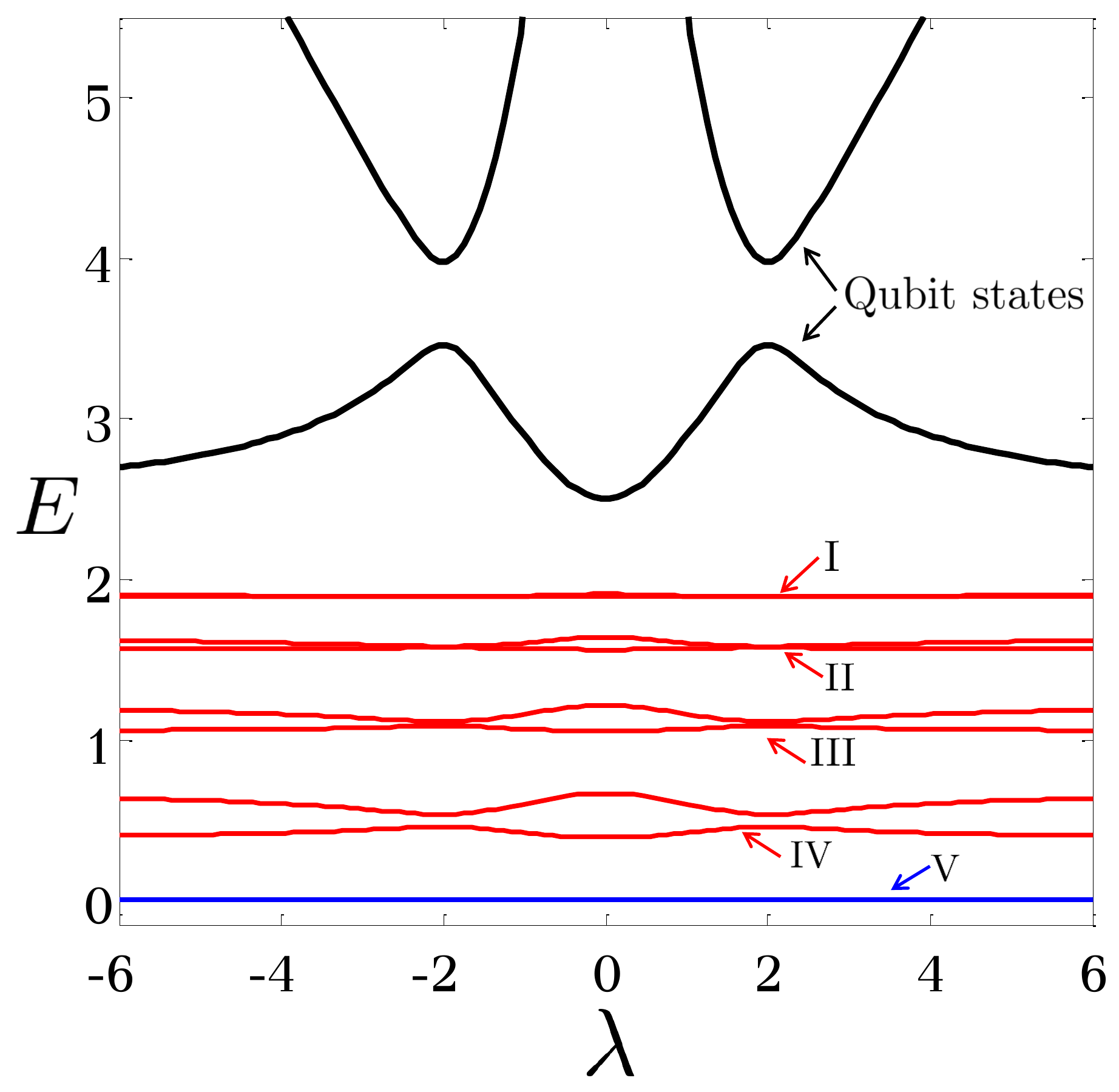}
\caption{Upper half of the band structure as a function of $\lambda$ for the open system with $\delta=2.5$, $\kappa=4$ and $\gamma=3$. The black curves correspond to states localized in the central qubits and the red curves are pairs of Bloch waves.}
\label{ComplexBandStructure}
\end{figure}

Next we consider the band structure of the open system. Fig.~\ref{ComplexBandStructure} shows the upper half of the band structure when $\gamma=3$. Here the $y$-axis gives the real part of energy. The picture is similar to the band structure of the closed system (Fig.~\ref{Bandstructure})
with all pairs slightly pushed towards the center of the band. In addition, pair~V has now become a superradiant pair where the two states have short lifetimes and are insensitive to the parameter $\lambda$. Therefore the pair is no longer a suitable candidate for our qubit system and its role is solely protecting other pairs from decaying into continuum.

\begin{figure}[h]
\centering
\captionsetup[subfigure]{}
\subfloat[]{
  \includegraphics[scale=0.35]{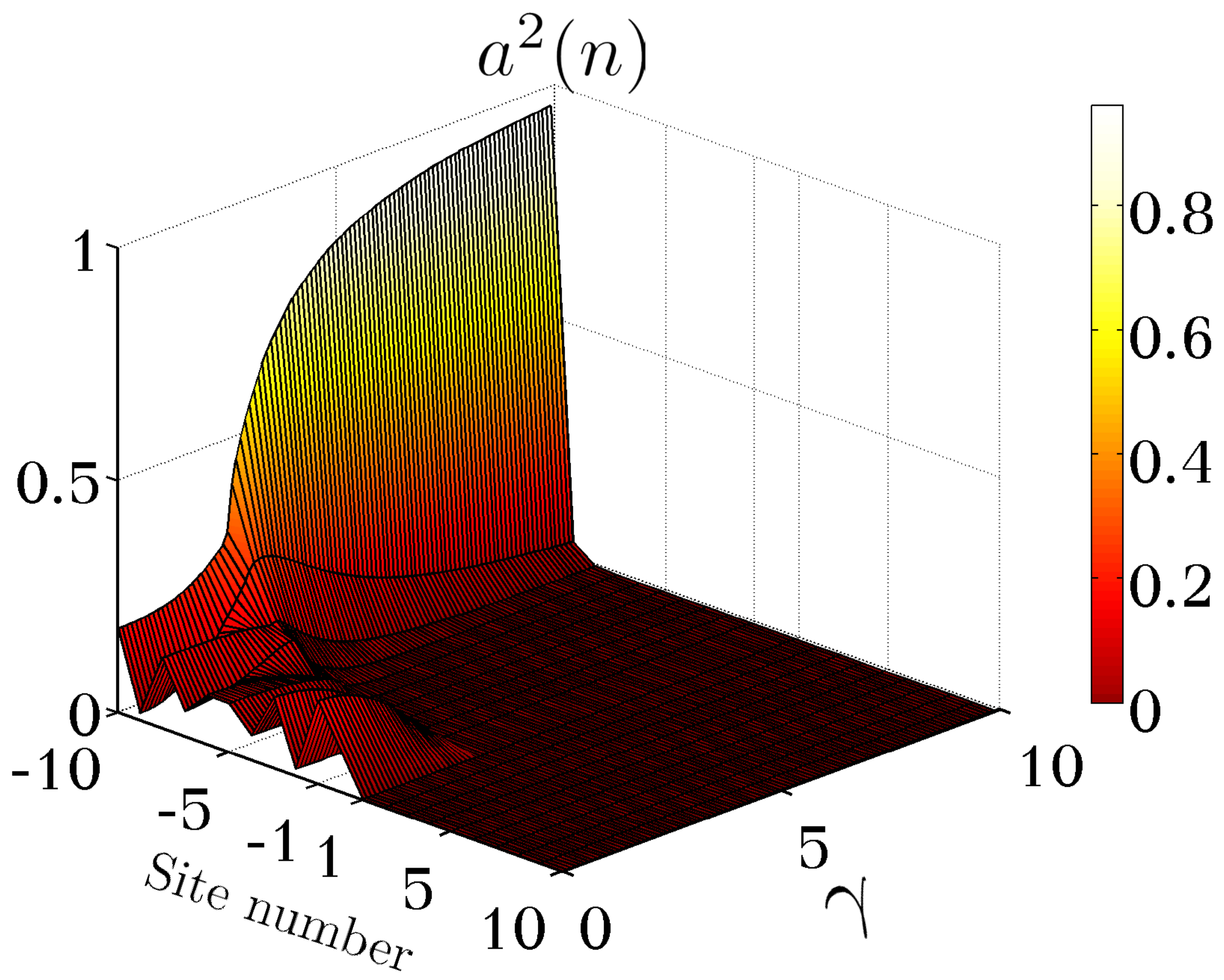}%
}
\captionsetup[subfigure]{}
\subfloat[]{%
  \includegraphics[scale=0.35]{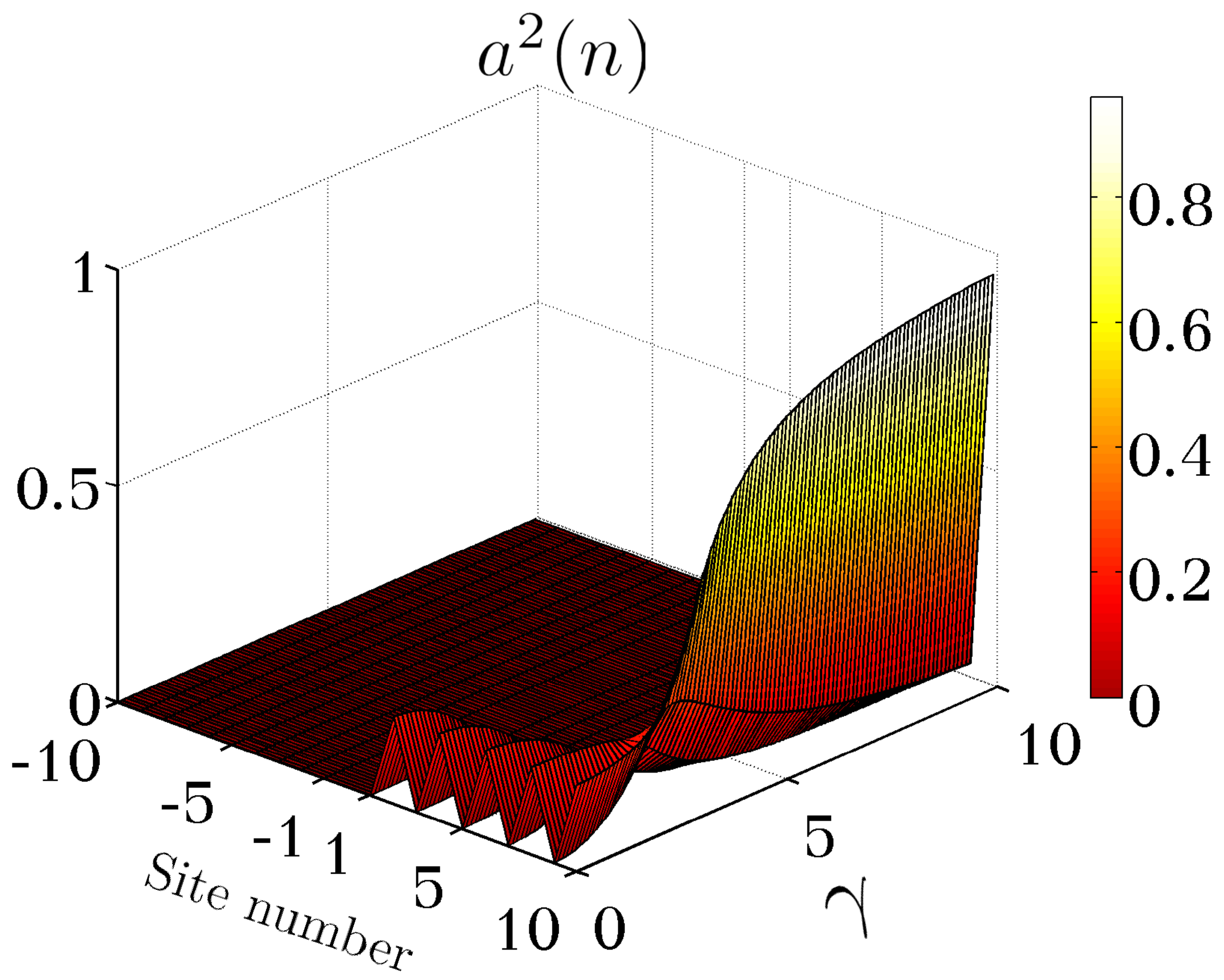}%
}
\captionsetup[subfigure]{justification=justified,singlelinecheck=false}
\caption{Evolution of the squared components of the wave functions for (a) upper and (b) lower superradiant states. The parameters are $\delta=2.5$, $\kappa=4$ and $\lambda=0.01$. Only the components inside the chain are considered in the figure.} \label{EdgeStatesLp01}
\end{figure}

It is interesting to monitor the superradiant wave function profile in the chain as a function of $\gamma$. We again consider two cases here: $\lambda=0.01$ and $\lambda=2$. The two superradiant wave functions for the first case are shown in Fig.~\ref{EdgeStatesLp01}. For small values of $\gamma$ the upper and lower states are Bloch waves confined to the left and the right sides of the chain, respectively. As $\gamma$ increases, the states quickly become localized at the  edges of the wire.

\begin{figure}[h]
\centering
\captionsetup[subfigure]{}
\subfloat[]{
  \includegraphics[scale=0.35]{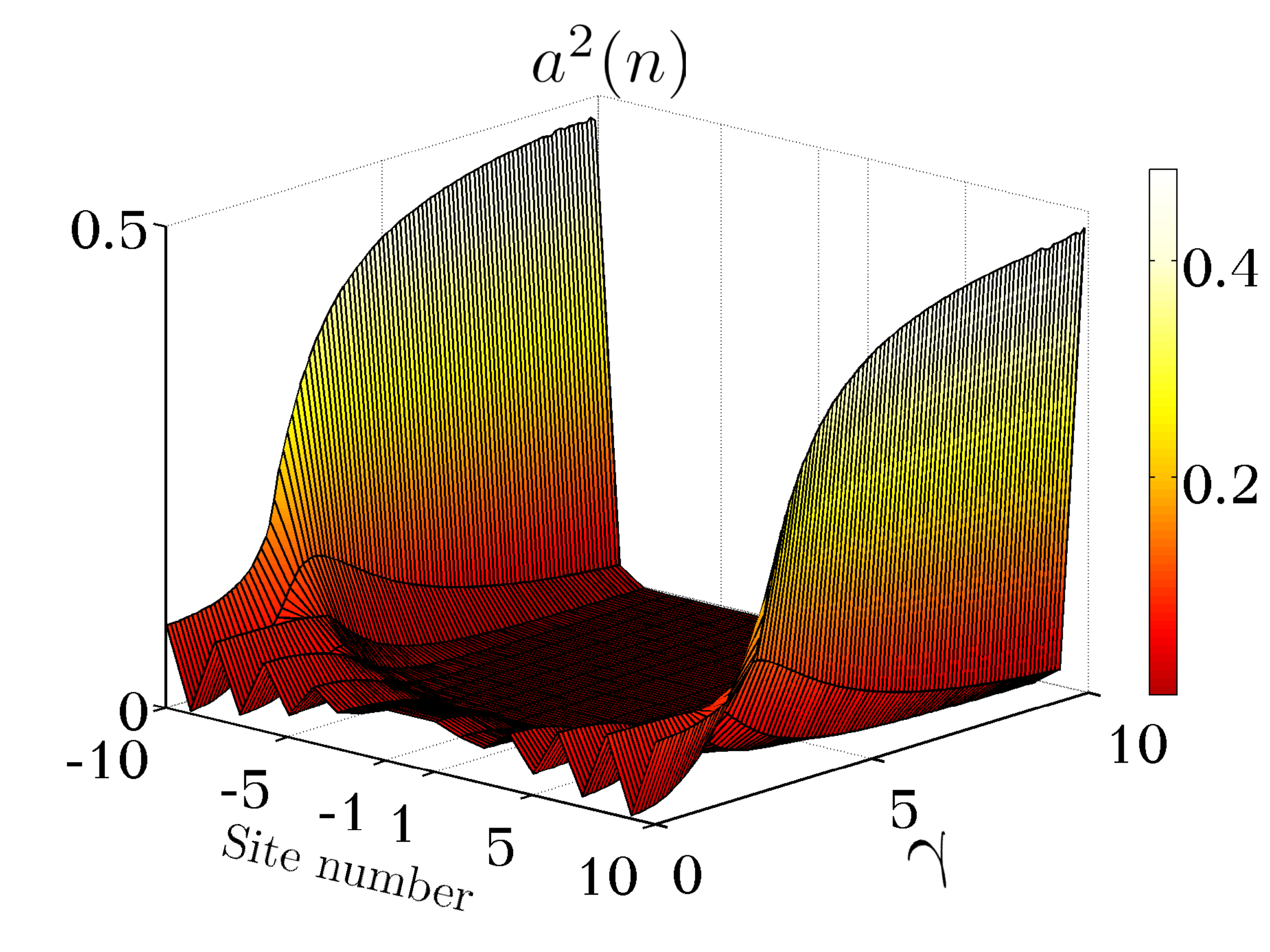}%
}
\captionsetup[subfigure]{}
\subfloat[]{%
  \includegraphics[scale=0.35]{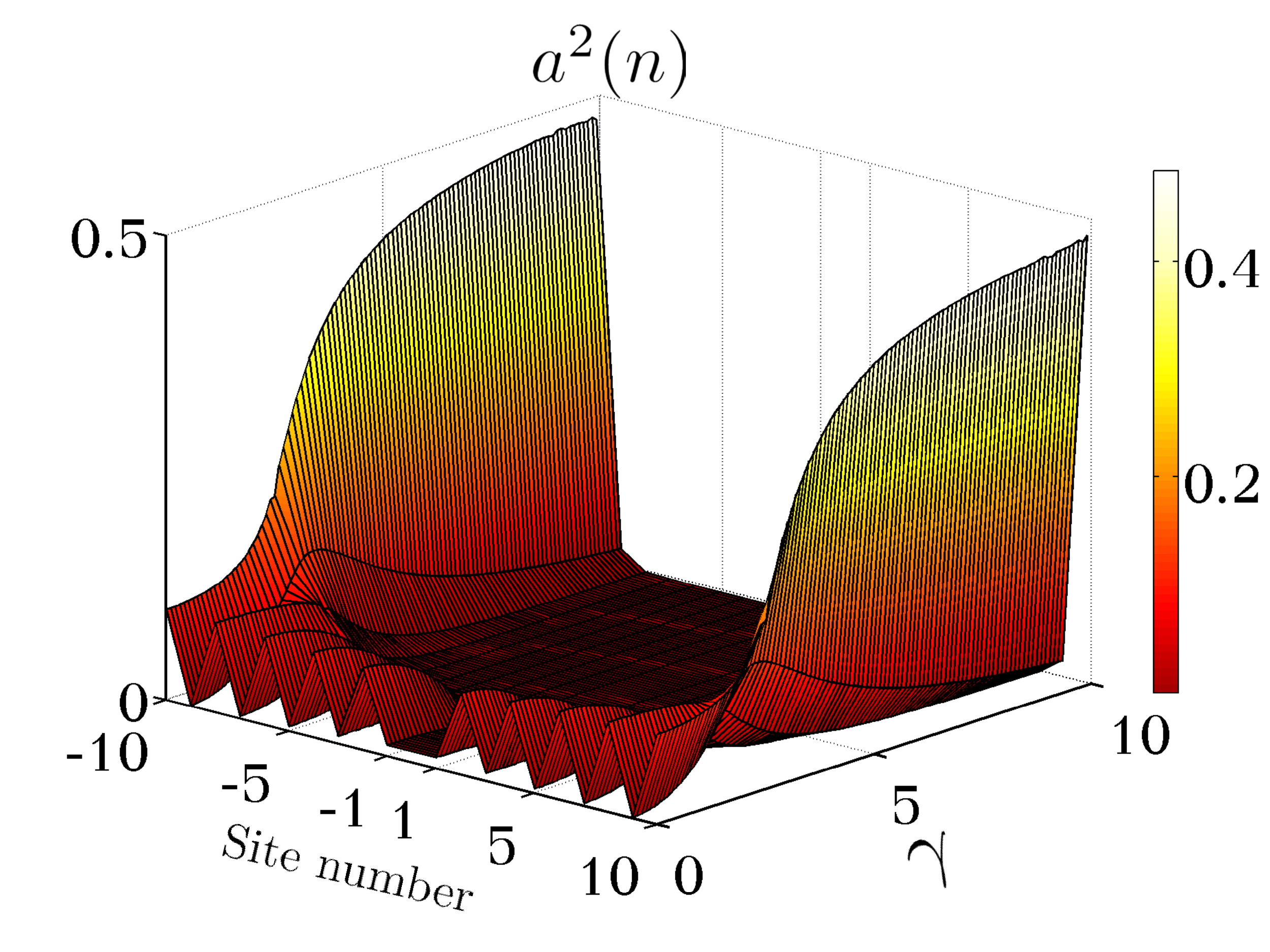}%
}
\captionsetup[subfigure]{justification=justified,singlelinecheck=false}
\caption{Evolution of the squared components of the wave functions for (a) upper and (b) lower superradiant states. The parameters are $\delta=2.5$, $\kappa=4$ and $\lambda=2$. Only the components inside the chain are considered in the figure.} \label{EdgeStatesL2}
\end{figure}

The wave functions for the second case, $\lambda=2$, are shown in Fig.~\ref{EdgeStatesL2}. The picture is similar to the previous case with the difference that, at weak coupling, the states are extended over the entire chain. It is apparent that at strong coupling the superradiant states become localized in the two edges \cite{volyazel}.

\section{Noise and decoherence} \label{secIV}

So far, we have studied the system under ideal conditions i.e. assuming that all processes are fully coherent. In reality, due to the interaction with the surrounding the system is subject to perturbations that might destroy the phase relations of the components of a wave function (decoherence). Because superradiance is a direct consequence of quantum superposition and therefore a coherent phenomenon, we expect such fluctuations to have a great impact on the dynamics of the system.

In this section we study the interplay between superradiance and decoherence by considering a model originally introduced by Haken and Strobl \cite{HakenStrobl}. The model has been extensively used for studying the role of dephasing in various physical situations such as radiative decay in molecular aggregates \cite{Grad}, quantum teleportation and the implementation of a quantum controlled-NOT gate \cite{Cheng}. The dephasing process is introduced by adding a time-dependent Hamiltonian to the effective non-Hermitian Hamiltonian,
\begin{equation}  \label{DephasingHamiltonian}
H_{\phi}(t)=\sum_{\substack{n=-N \\ n\neq 0}}^{N} \delta\epsilon_n(t) c^{\dagger}_{n}c_{n} + \delta\epsilon_L(t)c^{\dagger}_{L}c_{L}+ \delta\epsilon_R(t)c^{\dagger}_{R}c_{R},
\end{equation}
where $\delta\epsilon(t)$ describe stochastic Gaussian processes representing rapid fluctuations of on-site electronic energies with zero mean and delta-function correlations in time
\begin{equation} \label{StochProperty1Text}
\langle \delta\epsilon_i(t)\rangle=0,
\end{equation}
\begin{equation} \label{StochProperty2Text}
\langle \delta\epsilon_i(t) \delta\epsilon_j(t') \rangle=\alpha_{\phi} \delta_{i,j}\delta(t-t').
\end{equation}
Here $\alpha_{\phi}$ is the parameter representing the dephasing strength; subscripts $i$ and $j$ run over cell numbers, $n=-N,...,-1,1,...,N$, as well as the excited states of the central qubits, $L$ and $R$. The symbol $\langle \ \rangle$ denotes averaging over the statistical ensemble.

The evolution of the density matrix $\rho(t)$ is governed by the stochastic Liouville equation. In the site representation we have (see the appendix for derivation)
\begin{equation} \label{eqmotionavgrhoSpicMEText}
\frac{\partial}{\partial t}\langle \rho(t) \rangle_{i,j}= -i \big[\mathcal{H}_{\textrm{eff}},\langle \rho(t) \rangle\big]_{i,j} - 2 \alpha_{\phi} (1-\delta_{i,j})\langle \rho(t) \rangle_{i,j}.
\end{equation}
Both superradiance and dephasing result in the decay of off-diagonal elements of the density matrix. 
As a measure of coherence, we define a new quantity, $\mathcal{R}(t)$, according to
\begin{equation} \label{cohmeas}
\mathcal{R}(t)=\sum_{i\neq j}\langle \rho(t) \rangle_{i,j}.
\end{equation}

Furthermore, we define the coherence time, $\tau_{\textrm{coh}}$, as the time duration for which $\mathcal{R}(\tau_{\textrm{coh}})=\mathcal{R}(0)/e$. Without coupling to the continuum, $\gamma=0$, we have $\mathcal{R}(t)=\mathcal{R}(0)e^{-\alpha_{\phi}t}$ and 
$\tau_{\textrm{coh}}=\alpha_{\phi}^{-1}$. For an open system, $\gamma\neq 0$, the Liouville equation (\ref{eqmotionavgrhoSpicMEText}) was numerically solved using the ordinary differential equations package in Matlab. Here we consider three cases, $\alpha_{\phi}=10^{-3}$, $\alpha_{\phi}=10^{-2}$ and $\alpha_{\phi}=10^{-1}$. For each case, the coherence time is calculated for systems with different numbers of sites. The initial density matrix in all calculations corresponds to the upper state in the highest energy pair inside the Bloch band (pair I in Fig.~\ref{ComplexBandStructure}) when $\lambda^2=\kappa$.

\begin{figure}[h]
\centering
\includegraphics[scale=0.33]{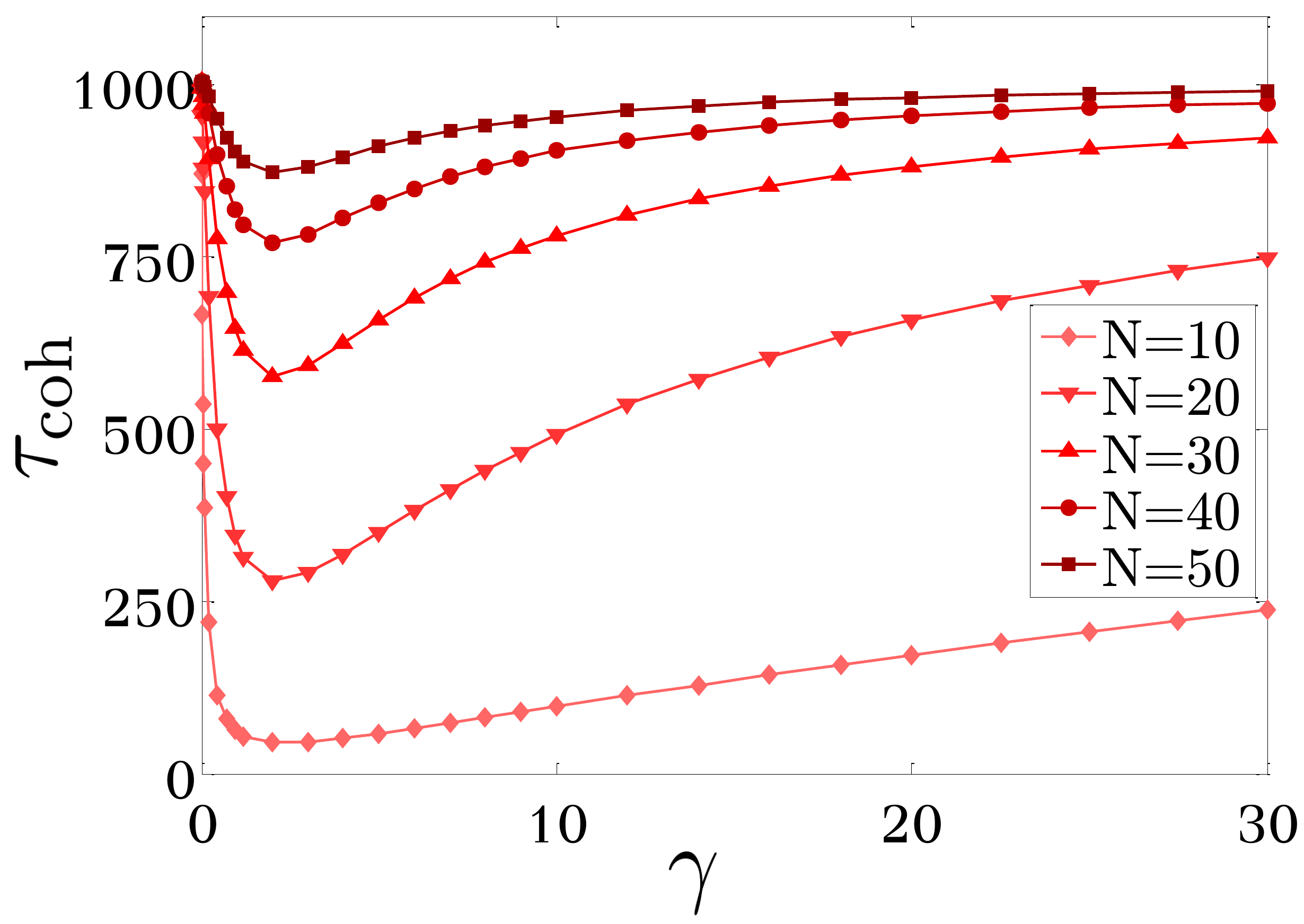}
\caption{Coherence time as defined in (\ref{cohmeas}) for systems with different numbers of cells when $\alpha_{\phi}=10^{-3}$. The initial state is the upper state of the highest pair inside the Bloch band. Other parameters are $\lambda=\sqrt{\kappa}=2$ and $\delta=2.5$.}
\label{GammaPhip001}
\end{figure}

The results for the case of the weak dephasing strength, $\alpha_{\phi}=10^{-3}$, are shown in Fig.~\ref{GammaPhip001}. As expected, regardless of the site number, the coherence time is equal to $\tau_{\textrm{coh}}=10^3$ when $\gamma=0$. As $\gamma$ increases, coherence time is governed by superradiance dynamics. At the transition to superradiance, the initial state achieves its maximum decay width and hence $\tau_{\textrm{coh}}$ reaches its minimum due to fast decay into the continuum. However, the reduction in $\tau_{\textrm{coh}}$ is quite different for chains with varying numbers of sites. For larger chains, states have a smaller share of the total width and therefore longer lifetimes compared to shorter chains.
Consequently, at the superradiance transition, longer chains have larger decoherence time. A similar type of resistance and robustness to noise with an increase in the number of sites was observed in nanoscale rings within light-harvesting systems \cite{CelardoR}. At larger values of $\gamma$, the lifetime of the initial state increases which in turn increases $\tau_{\textrm{coh}}$.

\begin{figure}[h]
\centering
\includegraphics[scale=0.33]{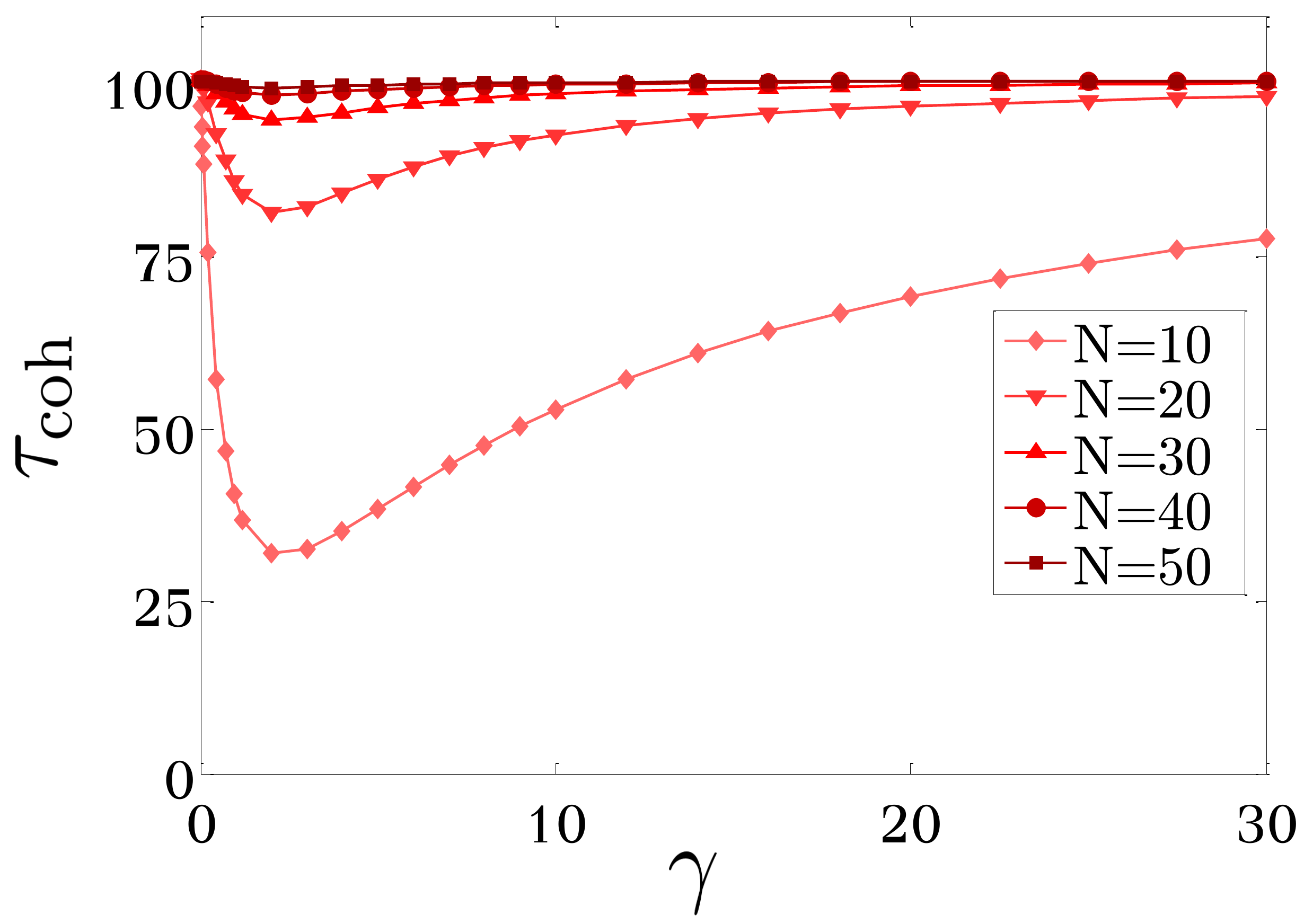}
\caption{Coherence time of systems with different numbers of cells when $\alpha_{\phi}=10^{-2}$. The initial state is the upper state of the highest pair inside the Bloch band.}
\label{GammaPhip01}
\end{figure}

Fig.~\ref{GammaPhip01} shows the results for $\alpha_{\phi}=10^{-2}$. The effect of superradiance is apparent only for systems with a smaller number of sites. For systems with $N=10$ and $N=20$, $\tau_{\textrm{coh}}$ is decreased due to the decay into the continuum (for smaller systems the lifetime of the initial state is shorter since there are less states that share the entire decay width).

\begin{figure}[h]
\centering
\includegraphics[scale=0.33]{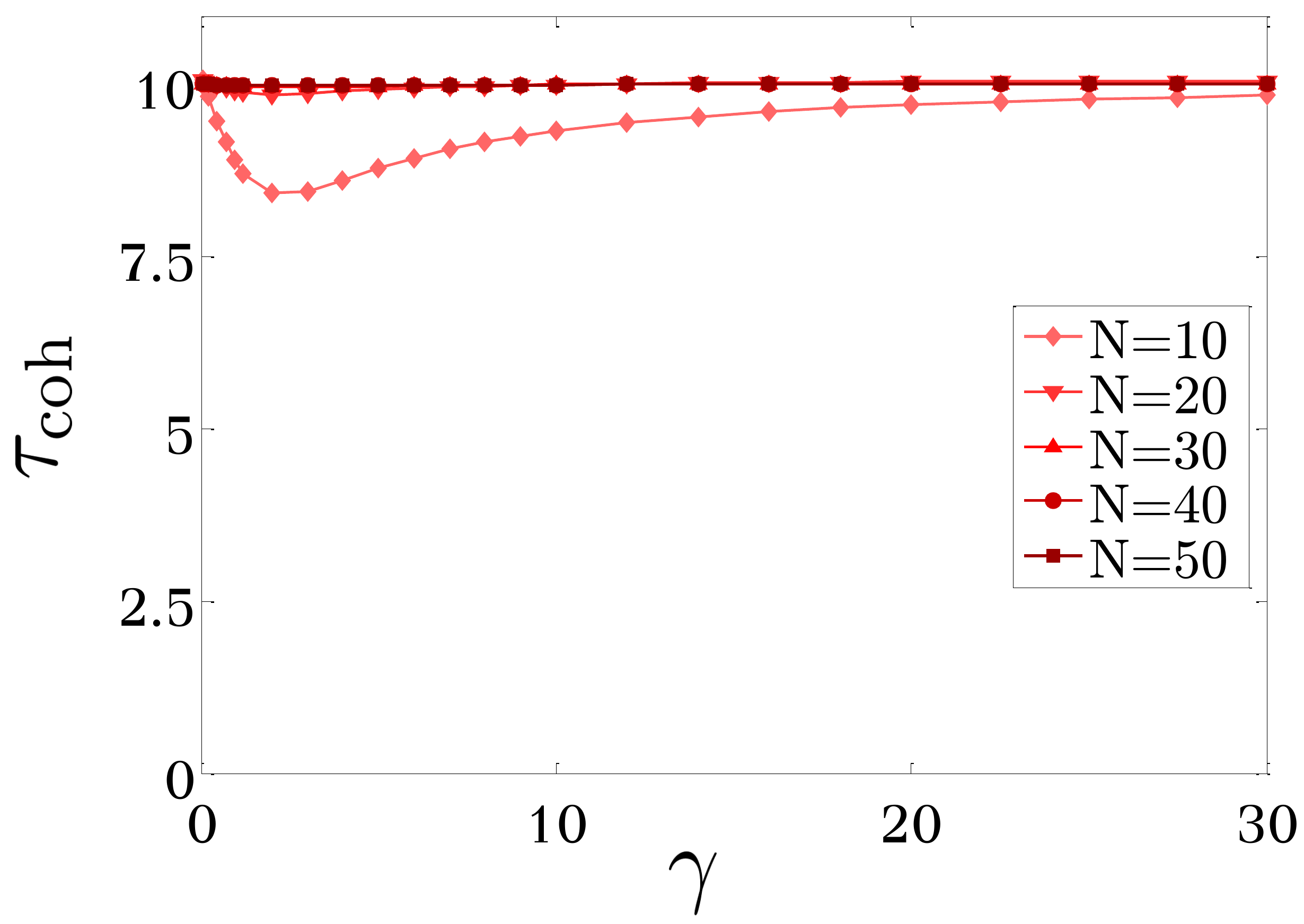}
\caption{Coherence time of systems with different numbers of cells when $\alpha_{\phi}=10^{-1}$. The initial state is the upper state of the highest pair inside the Bloch band.}
\label{GammaPhip1}
\end{figure}

The results associated with the case of 
$\alpha_{\phi}=10^{-1}$, are presented in Fig.~\ref{GammaPhip1}. The strong dephasing quickly dissipates the off-diagonal elements of the density matrix. This happens before the particle gets a chance to escape the wire and therefore the superradiance effect is suppressed by the dephasing phenomenon. As expected and demonstrated by the figures, superradiance survives only in the presence of relatively weak dephasing.

\section{Conclusion} \label{secV}

We proposed a novel solid-state based system for implementing a qubit. The system consists of a one-dimensional chain and a pair of two-level atoms inserted at its center where the couplings between their ground and excited states are inversely proportional. 
The energy eigenstates within the Bloch band 
exhibit a qubit-like behavior. The coupling between the ground and excited states of the two central qubits can be used to perform gate operations and initialize the system in a given state. The effect of connecting the wire to ideal leads and charge detectors (coupling to the continuum) was discussed in detail by exploiting the non-Hermitian effective Hamiltonian approach. In the case of strong continuum coupling, two broad resonances localized at the edges of the wire are formed. These emerging superradiant edge states increase the lifetime of the remaining states making them suitable candidates for qubit implementation. The effect of ambient noise was included by considering the stochastic Liouville equation. The interplay between superradiance and decoherence was discussed for various dephasing strengths. Superradiant effects are prominent for weak dephasing strengths and fade away as the noise increases.

\begin{acknowledgments}
A.~T. thanks A.~Stain for her support and helpful assistance. T.~H. and J.~W. thank Michigan State University and Dr.~G.~Richmond for holding and organizing the high school honors science, mathematics and engineering program (HSHSP). V.~Z. acknowledges support from the NSF grant PHY-1404442.
\end{acknowledgments}

\appendix

\section{The stochastic Liouville equation}

Here we provide a simple derivation of the stochastic Liouville equation using perturbation theory. The case considered here is a special case of the more general 
model where
the presence of phonons was accounted for by a heat bath affecting the electrons in a stochastic fashion. Similarly, we include the vibrational effects, which lead to dephasing, by the addition of the  time-dependent Hamiltonian (\ref{DephasingHamiltonian}). Fluctuations of electronic energies are modeled using Gaussian processes with 
the standard stochastic properties 
(\ref{StochProperty1Text}) and (\ref{StochProperty2Text}). The total Hamiltonian is then $H=\mathcal{H}+H_{\phi}$ and the evolution of the density operator is governed by the von Neumann equation,
\begin{equation} \label{vonNeumanneq}
\dot{\rho}= -i[H,\rho].
\end{equation}
Next we define superoperators $\mathcal{L}_{\textrm{eff}}$ and $\mathcal{L}_{\phi}$ according to
\begin{align} \label{defineL}
\mathcal{L}_{\textrm{eff}} \rho &= [\mathcal{H},\rho], \\
\mathcal{L}_{\phi} \rho &= [H_{\phi},\rho] .
\end{align}
The von Neumann equation (\ref{vonNeumanneq}) in terms of the superoperators reads
\begin{equation} \label{vonNeumanneq2}
\dot{\rho}= -i\mathcal{L}\rho=-i(\mathcal{L}_{\textrm{eff}}+\mathcal{L}_{\phi})\rho.
\end{equation}
Using
\begin{align}
\rho^{I}(t) &= e^{i\mathcal{L}_{\textrm{eff}}t}\rho(t), \\
\mathcal{L}_{\phi}^{I}  &= e^{i\mathcal{L}_{\textrm{eff}}t} \mathcal{L}_{\phi}  e^{-i\mathcal{L}_{\textrm{eff}}t},
\end{align}
the time evolution equation, (\ref{vonNeumanneq2}), transforms to the interaction picture,
\begin{equation} \label{vonNeumanneqInt}
\frac{\partial}{\partial t} \rho^{I}(t)= -i\mathcal{L}_{\phi}^{I} \rho^{I}.
\end{equation}
The solution of eq.~(\ref{vonNeumanneqInt}) up to the second order in the perturbation expansion is
\begin{align} \label{StochEqnPT}
\rho^{I}(t) =& \rho^{I}(0)-i\int^{t}_{0} dt' \mathcal{L}_{\phi}^{I}(t') \rho^{I}(0) \nonumber \\
&+i^2 \int_{0}^{t} dt' \int_{0}^{t'} dt'' \mathcal{L}_{\phi}^{I}(t')\mathcal{L}_{\phi}^{I}(t'') \rho^{I}(0).
\end{align}

This formal solution only makes sense after averaging over the ensemble. Using the properties of the bath given in eqs.~(\ref{StochProperty1Text}) and (\ref{StochProperty2Text}) we have
\begin{equation} \label{StochEqnPTavg}
\langle \rho^{I}(t) \rangle= \rho^{I}(0)-\int_{0}^{t} dt' \int_{0}^{t'} dt'' \langle \mathcal{L}_{\phi}^{I}(t')\mathcal{L}_{\phi}^{I}(t'') \rangle \rho^{I}(0).
\end{equation}

Differentiating (\ref{StochEqnPTavg}) 
we arrive at the equation of motion for $\langle \rho^{I}(t) \rangle$,
\begin{equation} \label{eqmotionavgrho}
\frac{\partial}{\partial t}\langle \rho^{I}(t) \rangle= -\int_{0}^{t} dt' \langle \mathcal{L}_{\phi}^{I}(t)\mathcal{L}_{\phi}^{I}(t') \rangle \langle \rho^{I}(t) \rangle.
\end{equation}
In obtaining (\ref{eqmotionavgrho}) it was assumed that the bath memory is short compared to the time scales of the density operator. Therefore $\rho^{I}(0)$ was replaced by $\langle \rho^{I}(t) \rangle$. Evaluating the integral with the help of (\ref{StochProperty2Text}) and using the definition provided in (\ref{defineL}) we arrive at
\begin{equation} \label{eqmotionavgrho2}
\frac{\partial}{\partial t}\langle \rho^{I}(t) \rangle= -\alpha_{\phi} \sum_{j} \Big[(c_{j}^I)^{\dagger}c_{j}^I,\big[(c_{j}^{I})^{\dagger}c_{j}^I,\langle \rho^{I}(t)\rangle \big]\Big],
\end{equation}
where $j$ runs over cells, $n=-N,...,-1,1,...,N$, as well as the excited states of the central qubits, $L$ and $R$. Going back to the Schr\"{o}dinger picture we have
\begin{equation} \label{eqmotionavgrhoSpic}
\frac{\partial}{\partial t}\langle \rho(t) \rangle= -i \mathcal{L}_{\textrm{eff}} \langle \rho(t) \rangle - \alpha_{\phi} \sum_{j} \Big[c_{j}^{\dagger}c_{j},\big[c_{j}^{\dagger}c_{j},\langle \rho(t)\rangle \big]\Big].
\end{equation}

Finally, by calculating the matrix element of the density operator in the site representation, we arrive at the expression given in (\ref{eqmotionavgrhoSpicMEText}),
\begin{equation} \label{eqmotionavgrhoSpicME}
\frac{\partial}{\partial t}\langle \rho(t) \rangle_{i,j}= -i \big[\mathcal{H}_{\textrm{eff}},\langle \rho(t) \rangle\big]_{i,j} - 2 \alpha_{\phi} (1-\delta_{i,j})\langle \rho(t) \rangle_{i,j}.
\end{equation}
Even though we have used perturbation expansion to derive the above stochastic Liouville equation, 
the final result is exact \cite{Reineker} due to the Markovian character of the random process, eqs. (32) and (33).

\bibliography{Refs}

\begin{thebibliography}{43}%
\makeatletter
\providecommand \@ifxundefined [1]{%
 \@ifx{#1\undefined}
}%
\providecommand \@ifnum [1]{%
 \ifnum #1\expandafter \@firstoftwo
 \else \expandafter \@secondoftwo
 \fi
}%
\providecommand \@ifx [1]{%
 \ifx #1\expandafter \@firstoftwo
 \else \expandafter \@secondoftwo
 \fi
}%
\providecommand \natexlab [1]{#1}%
\providecommand \enquote  [1]{``#1''}%
\providecommand \bibnamefont  [1]{#1}%
\providecommand \bibfnamefont [1]{#1}%
\providecommand \citenamefont [1]{#1}%
\providecommand \href@noop [0]{\@secondoftwo}%
\providecommand \href [0]{\begingroup \@sanitize@url \@href}%
\providecommand \@href[1]{\@@startlink{#1}\@@href}%
\providecommand \@@href[1]{\endgroup#1\@@endlink}%
\providecommand \@sanitize@url [0]{\catcode `\\12\catcode `\$12\catcode
  `\&12\catcode `\#12\catcode `\^12\catcode `\_12\catcode `\%12\relax}%
\providecommand \@@startlink[1]{}%
\providecommand \@@endlink[0]{}%
\providecommand \url  [0]{\begingroup\@sanitize@url \@url }%
\providecommand \@url [1]{\endgroup\@href {#1}{\urlprefix }}%
\providecommand \urlprefix  [0]{URL }%
\providecommand \Eprint [0]{\href }%
\providecommand \doibase [0]{http://dx.doi.org/}%
\providecommand \selectlanguage [0]{\@gobble}%
\providecommand \bibinfo  [0]{\@secondoftwo}%
\providecommand \bibfield  [0]{\@secondoftwo}%
\providecommand \translation [1]{[#1]}%
\providecommand \BibitemOpen [0]{}%
\providecommand \bibitemStop [0]{}%
\providecommand \bibitemNoStop [0]{.\EOS\space}%
\providecommand \EOS [0]{\spacefactor3000\relax}%
\providecommand \BibitemShut  [1]{\csname bibitem#1\endcsname}%
\let\auto@bib@innerbib\@empty
\bibitem [{\citenamefont {O{\textquoteright}Brien}(2007)}]{opticQC1}%
  \BibitemOpen
  \bibfield  {author} {\bibinfo {author} {\bibfnamefont {J.~L.}\ \bibnamefont
  {O{\textquoteright}Brien}},\ }\href {\doibase 10.1126/science.1142892}
  {\bibfield  {journal} {\bibinfo  {journal} {Science}\ }\textbf {\bibinfo
  {volume} {318}},\ \bibinfo {pages} {1567} (\bibinfo {year}
  {2007})}\BibitemShut {NoStop}%
\bibitem [{\citenamefont {Kok}\ \emph {et~al.}(2007)\citenamefont {Kok},
  \citenamefont {Munro}, \citenamefont {Nemoto}, \citenamefont {Ralph},
  \citenamefont {Dowling},\ and\ \citenamefont {Milburn}}]{opticQC2}%
  \BibitemOpen
  \bibfield  {author} {\bibinfo {author} {\bibfnamefont {P.}~\bibnamefont
  {Kok}}, \bibinfo {author} {\bibfnamefont {W.~J.}\ \bibnamefont {Munro}},
  \bibinfo {author} {\bibfnamefont {K.}~\bibnamefont {Nemoto}}, \bibinfo
  {author} {\bibfnamefont {T.~C.}\ \bibnamefont {Ralph}}, \bibinfo {author}
  {\bibfnamefont {J.~P.}\ \bibnamefont {Dowling}}, \ and\ \bibinfo {author}
  {\bibfnamefont {G.~J.}\ \bibnamefont {Milburn}},\ }\href {\doibase
  10.1103/RevModPhys.79.135} {\bibfield  {journal} {\bibinfo  {journal} {Rev.
  Mod. Phys.}\ }\textbf {\bibinfo {volume} {79}},\ \bibinfo {pages} {135}
  (\bibinfo {year} {2007})}\BibitemShut {NoStop}%
\bibitem [{\citenamefont {Warren}(1997)}]{NMRQC1}%
  \BibitemOpen
  \bibfield  {author} {\bibinfo {author} {\bibfnamefont {W.~S.}\ \bibnamefont
  {Warren}},\ }\href {\doibase 10.1126/science.277.5332.1688} {\bibfield
  {journal} {\bibinfo  {journal} {Science}\ }\textbf {\bibinfo {volume}
  {277}},\ \bibinfo {pages} {1688} (\bibinfo {year} {1997})}\BibitemShut
  {NoStop}%
\bibitem [{\citenamefont {Mamin}\ \emph {et~al.}(2013)\citenamefont {Mamin},
  \citenamefont {Kim}, \citenamefont {Sherwood}, \citenamefont {Rettner},
  \citenamefont {Ohno}, \citenamefont {Awschalom},\ and\ \citenamefont
  {Rugar}}]{NMRQC2}%
  \BibitemOpen
  \bibfield  {author} {\bibinfo {author} {\bibfnamefont {H.~J.}\ \bibnamefont
  {Mamin}}, \bibinfo {author} {\bibfnamefont {M.}~\bibnamefont {Kim}}, \bibinfo
  {author} {\bibfnamefont {M.~H.}\ \bibnamefont {Sherwood}}, \bibinfo {author}
  {\bibfnamefont {C.~T.}\ \bibnamefont {Rettner}}, \bibinfo {author}
  {\bibfnamefont {K.}~\bibnamefont {Ohno}}, \bibinfo {author} {\bibfnamefont
  {D.~D.}\ \bibnamefont {Awschalom}}, \ and\ \bibinfo {author} {\bibfnamefont
  {D.}~\bibnamefont {Rugar}},\ }\href {\doibase 10.1126/science.1231540}
  {\bibfield  {journal} {\bibinfo  {journal} {Science}\ }\textbf {\bibinfo
  {volume} {339}},\ \bibinfo {pages} {557} (\bibinfo {year}
  {2013})}\BibitemShut {NoStop}%
\bibitem [{\citenamefont {Xu}\ \emph {et~al.}(2012)\citenamefont {Xu},
  \citenamefont {Zhu}, \citenamefont {Lu}, \citenamefont {Zhou}, \citenamefont
  {Peng},\ and\ \citenamefont {Du}}]{NMRQC3}%
  \BibitemOpen
  \bibfield  {author} {\bibinfo {author} {\bibfnamefont {N.}~\bibnamefont
  {Xu}}, \bibinfo {author} {\bibfnamefont {J.}~\bibnamefont {Zhu}}, \bibinfo
  {author} {\bibfnamefont {D.}~\bibnamefont {Lu}}, \bibinfo {author}
  {\bibfnamefont {X.}~\bibnamefont {Zhou}}, \bibinfo {author} {\bibfnamefont
  {X.}~\bibnamefont {Peng}}, \ and\ \bibinfo {author} {\bibfnamefont
  {J.}~\bibnamefont {Du}},\ }\href {\doibase 10.1103/PhysRevLett.108.130501}
  {\bibfield  {journal} {\bibinfo  {journal} {Phys. Rev. Lett.}\ }\textbf
  {\bibinfo {volume} {108}},\ \bibinfo {pages} {130501} (\bibinfo {year}
  {2012})}\BibitemShut {NoStop}%
\bibitem [{\citenamefont {H{\"a}ffner}\ \emph {et~al.}(2008)\citenamefont
  {H{\"a}ffner}, \citenamefont {Roos},\ and\ \citenamefont
  {Blatt}}]{TrapIonQC1}%
  \BibitemOpen
  \bibfield  {author} {\bibinfo {author} {\bibfnamefont {H.}~\bibnamefont
  {H{\"a}ffner}}, \bibinfo {author} {\bibfnamefont {C.}~\bibnamefont {Roos}}, \
  and\ \bibinfo {author} {\bibfnamefont {R.}~\bibnamefont {Blatt}},\ }\href
  {\doibase http://dx.doi.org/10.1016/j.physrep.2008.09.003} {\bibfield
  {journal} {\bibinfo  {journal} {Phys. Rep.}\ }\textbf {\bibinfo {volume}
  {469}},\ \bibinfo {pages} {155 } (\bibinfo {year} {2008})}\BibitemShut
  {NoStop}%
\bibitem [{\citenamefont {Kielpinski}\ \emph {et~al.}(2002)\citenamefont
  {Kielpinski}, \citenamefont {Monroe},\ and\ \citenamefont
  {Wineland}}]{TrapIonQC2}%
  \BibitemOpen
  \bibfield  {author} {\bibinfo {author} {\bibfnamefont {D.}~\bibnamefont
  {Kielpinski}}, \bibinfo {author} {\bibfnamefont {C.}~\bibnamefont {Monroe}},
  \ and\ \bibinfo {author} {\bibfnamefont {D.~J.}\ \bibnamefont {Wineland}},\
  }\href@noop {} {\bibfield  {journal} {\bibinfo  {journal} {Nature}\ }\textbf
  {\bibinfo {volume} {417}},\ \bibinfo {pages} {709} (\bibinfo {year}
  {2002})}\BibitemShut {NoStop}%
\bibitem [{\citenamefont {Monroe}\ and\ \citenamefont
  {Kim}(2013)}]{TrapIonQC3}%
  \BibitemOpen
  \bibfield  {author} {\bibinfo {author} {\bibfnamefont {C.}~\bibnamefont
  {Monroe}}\ and\ \bibinfo {author} {\bibfnamefont {J.}~\bibnamefont {Kim}},\
  }\href {\doibase 10.1126/science.1231298} {\bibfield  {journal} {\bibinfo
  {journal} {Science}\ }\textbf {\bibinfo {volume} {339}},\ \bibinfo {pages}
  {1164} (\bibinfo {year} {2013})}\BibitemShut {NoStop}%
\bibitem [{\citenamefont {Devoret}\ and\ \citenamefont
  {Schoelkopf}(2013)}]{SPCQ1}%
  \BibitemOpen
  \bibfield  {author} {\bibinfo {author} {\bibfnamefont {M.~H.}\ \bibnamefont
  {Devoret}}\ and\ \bibinfo {author} {\bibfnamefont {R.~J.}\ \bibnamefont
  {Schoelkopf}},\ }\href {\doibase 10.1126/science.1231930} {\bibfield
  {journal} {\bibinfo  {journal} {Science}\ }\textbf {\bibinfo {volume}
  {339}},\ \bibinfo {pages} {1169} (\bibinfo {year} {2013})}\BibitemShut
  {NoStop}%
\bibitem [{\citenamefont {Flurin}\ \emph {et~al.}(2015)\citenamefont {Flurin},
  \citenamefont {Roch}, \citenamefont {Pillet}, \citenamefont {Mallet},\ and\
  \citenamefont {Huard}}]{SPCQ2}%
  \BibitemOpen
  \bibfield  {author} {\bibinfo {author} {\bibfnamefont {E.}~\bibnamefont
  {Flurin}}, \bibinfo {author} {\bibfnamefont {N.}~\bibnamefont {Roch}},
  \bibinfo {author} {\bibfnamefont {J.~D.}\ \bibnamefont {Pillet}}, \bibinfo
  {author} {\bibfnamefont {F.}~\bibnamefont {Mallet}}, \ and\ \bibinfo {author}
  {\bibfnamefont {B.}~\bibnamefont {Huard}},\ }\href {\doibase
  10.1103/PhysRevLett.114.090503} {\bibfield  {journal} {\bibinfo  {journal}
  {Phys. Rev. Lett.}\ }\textbf {\bibinfo {volume} {114}},\ \bibinfo {pages}
  {090503} (\bibinfo {year} {2015})}\BibitemShut {NoStop}%
\bibitem [{\citenamefont {Douce}\ \emph {et~al.}(2015)\citenamefont {Douce},
  \citenamefont {Stern}, \citenamefont {Zagury}, \citenamefont {Bertet},\ and\
  \citenamefont {Milman}}]{SPCQ3}%
  \BibitemOpen
  \bibfield  {author} {\bibinfo {author} {\bibfnamefont {T.}~\bibnamefont
  {Douce}}, \bibinfo {author} {\bibfnamefont {M.}~\bibnamefont {Stern}},
  \bibinfo {author} {\bibfnamefont {N.}~\bibnamefont {Zagury}}, \bibinfo
  {author} {\bibfnamefont {P.}~\bibnamefont {Bertet}}, \ and\ \bibinfo {author}
  {\bibfnamefont {P.}~\bibnamefont {Milman}},\ }\href {\doibase
  10.1103/PhysRevA.92.052335} {\bibfield  {journal} {\bibinfo  {journal} {Phys.
  Rev. A}\ }\textbf {\bibinfo {volume} {92}},\ \bibinfo {pages} {052335}
  (\bibinfo {year} {2015})}\BibitemShut {NoStop}%
\bibitem [{\citenamefont {Lucero}\ \emph {et~al.}(2012)\citenamefont {Lucero},
  \citenamefont {Barends}, \citenamefont {Chen}, \citenamefont {Kelly},
  \citenamefont {Mariantoni}, \citenamefont {Megrant}, \citenamefont
  {O'Malley}, \citenamefont {Sank}, \citenamefont {Vainsencher}, \citenamefont
  {Wenner}, \citenamefont {White}, \citenamefont {Yin}, \citenamefont
  {Cleland},\ and\ \citenamefont {Martinis}}]{SPCQ4}%
  \BibitemOpen
  \bibfield  {author} {\bibinfo {author} {\bibfnamefont {E.}~\bibnamefont
  {Lucero}}, \bibinfo {author} {\bibfnamefont {R.}~\bibnamefont {Barends}},
  \bibinfo {author} {\bibfnamefont {Y.}~\bibnamefont {Chen}}, \bibinfo {author}
  {\bibfnamefont {J.}~\bibnamefont {Kelly}}, \bibinfo {author} {\bibfnamefont
  {M.}~\bibnamefont {Mariantoni}}, \bibinfo {author} {\bibfnamefont
  {A.}~\bibnamefont {Megrant}}, \bibinfo {author} {\bibfnamefont
  {P.}~\bibnamefont {O'Malley}}, \bibinfo {author} {\bibfnamefont
  {D.}~\bibnamefont {Sank}}, \bibinfo {author} {\bibfnamefont {A.}~\bibnamefont
  {Vainsencher}}, \bibinfo {author} {\bibfnamefont {J.}~\bibnamefont {Wenner}},
  \bibinfo {author} {\bibfnamefont {T.}~\bibnamefont {White}}, \bibinfo
  {author} {\bibfnamefont {Y.}~\bibnamefont {Yin}}, \bibinfo {author}
  {\bibfnamefont {A.~N.}\ \bibnamefont {Cleland}}, \ and\ \bibinfo {author}
  {\bibfnamefont {J.~M.}\ \bibnamefont {Martinis}},\ }\href@noop {} {\bibfield
  {journal} {\bibinfo  {journal} {Nat. Phys.}\ }\textbf {\bibinfo {volume}
  {8}},\ \bibinfo {pages} {719} (\bibinfo {year} {2012})}\BibitemShut {NoStop}%
\bibitem [{\citenamefont {Platzman}\ and\ \citenamefont
  {Dykman}(1999)}]{Dykman0}%
  \BibitemOpen
  \bibfield  {author} {\bibinfo {author} {\bibfnamefont {P.~M.}\ \bibnamefont
  {Platzman}}\ and\ \bibinfo {author} {\bibfnamefont {M.~I.}\ \bibnamefont
  {Dykman}},\ }\href {\doibase 10.1126/science.284.5422.1967} {\bibfield
  {journal} {\bibinfo  {journal} {Science}\ }\textbf {\bibinfo {volume}
  {284}},\ \bibinfo {pages} {1967} (\bibinfo {year} {1999})}\BibitemShut
  {NoStop}%
\bibitem [{\citenamefont {Dykman}\ \emph {et~al.}(2003)\citenamefont {Dykman},
  \citenamefont {Platzman},\ and\ \citenamefont {Seddighrad}}]{Dykman}%
  \BibitemOpen
  \bibfield  {author} {\bibinfo {author} {\bibfnamefont {M.~I.}\ \bibnamefont
  {Dykman}}, \bibinfo {author} {\bibfnamefont {P.~M.}\ \bibnamefont
  {Platzman}}, \ and\ \bibinfo {author} {\bibfnamefont {P.}~\bibnamefont
  {Seddighrad}},\ }\href {\doibase 10.1103/PhysRevB.67.155402} {\bibfield
  {journal} {\bibinfo  {journal} {Phys. Rev. B}\ }\textbf {\bibinfo {volume}
  {67}},\ \bibinfo {pages} {155402} (\bibinfo {year} {2003})}\BibitemShut
  {NoStop}%
\bibitem [{\citenamefont {Yamamoto}\ \emph {et~al.}(2012)\citenamefont
  {Yamamoto}, \citenamefont {Takada}, \citenamefont {Bauerle}, \citenamefont
  {Watanabe}, \citenamefont {Wieck},\ and\ \citenamefont {Tarucha}}]{SSQC1}%
  \BibitemOpen
  \bibfield  {author} {\bibinfo {author} {\bibfnamefont {M.}~\bibnamefont
  {Yamamoto}}, \bibinfo {author} {\bibfnamefont {S.}~\bibnamefont {Takada}},
  \bibinfo {author} {\bibfnamefont {C.}~\bibnamefont {Bauerle}}, \bibinfo
  {author} {\bibfnamefont {K.}~\bibnamefont {Watanabe}}, \bibinfo {author}
  {\bibfnamefont {A.~D.}\ \bibnamefont {Wieck}}, \ and\ \bibinfo {author}
  {\bibfnamefont {S.}~\bibnamefont {Tarucha}},\ }\href@noop {} {\bibfield
  {journal} {\bibinfo  {journal} {Nat. Nano.}\ }\textbf {\bibinfo {volume}
  {7}},\ \bibinfo {pages} {247} (\bibinfo {year} {2012})}\BibitemShut {NoStop}%
\bibitem [{\citenamefont {Petersson}\ \emph {et~al.}(2010)\citenamefont
  {Petersson}, \citenamefont {Petta}, \citenamefont {Lu},\ and\ \citenamefont
  {Gossard}}]{SSQC2}%
  \BibitemOpen
  \bibfield  {author} {\bibinfo {author} {\bibfnamefont {K.~D.}\ \bibnamefont
  {Petersson}}, \bibinfo {author} {\bibfnamefont {J.~R.}\ \bibnamefont
  {Petta}}, \bibinfo {author} {\bibfnamefont {H.}~\bibnamefont {Lu}}, \ and\
  \bibinfo {author} {\bibfnamefont {A.~C.}\ \bibnamefont {Gossard}},\ }\href
  {\doibase 10.1103/PhysRevLett.105.246804} {\bibfield  {journal} {\bibinfo
  {journal} {Phys. Rev. Lett.}\ }\textbf {\bibinfo {volume} {105}},\ \bibinfo
  {pages} {246804} (\bibinfo {year} {2010})}\BibitemShut {NoStop}%
\bibitem [{\citenamefont {Hayashi}\ \emph {et~al.}(2003)\citenamefont
  {Hayashi}, \citenamefont {Fujisawa}, \citenamefont {Cheong}, \citenamefont
  {Jeong},\ and\ \citenamefont {Hirayama}}]{SSQC3}%
  \BibitemOpen
  \bibfield  {author} {\bibinfo {author} {\bibfnamefont {T.}~\bibnamefont
  {Hayashi}}, \bibinfo {author} {\bibfnamefont {T.}~\bibnamefont {Fujisawa}},
  \bibinfo {author} {\bibfnamefont {H.~D.}\ \bibnamefont {Cheong}}, \bibinfo
  {author} {\bibfnamefont {Y.~H.}\ \bibnamefont {Jeong}}, \ and\ \bibinfo
  {author} {\bibfnamefont {Y.}~\bibnamefont {Hirayama}},\ }\href {\doibase
  10.1103/PhysRevLett.91.226804} {\bibfield  {journal} {\bibinfo  {journal}
  {Phys. Rev. Lett.}\ }\textbf {\bibinfo {volume} {91}},\ \bibinfo {pages}
  {226804} (\bibinfo {year} {2003})}\BibitemShut {NoStop}%
\bibitem [{\citenamefont {Pavi{\v{c}}i{\'c}}(2013)}]{pavivcic}%
  \BibitemOpen
  \bibfield  {author} {\bibinfo {author} {\bibfnamefont {M.}~\bibnamefont
  {Pavi{\v{c}}i{\'c}}},\ }\href@noop {} {\emph {\bibinfo {title} {Companion to
  quantum computation and communication}}}\ (\bibinfo  {publisher} {Wiley-VCH,
  Darmstadt},\ \bibinfo {year} {2013})\BibitemShut {NoStop}%
\bibitem [{\citenamefont {Ardavan}\ \emph {et~al.}(2003)\citenamefont
  {Ardavan}, \citenamefont {Austwick}, \citenamefont {Benjamin}, \citenamefont
  {Briggs}, \citenamefont {Denni~s}, \citenamefont {Ferguson}, \citenamefont
  {Hasko}, \citenamefont {Kanai}, \citenamefont {Khlobystov}, \citenamefont
  {Lovett}, \citenamefont {Morley}, \citenamefont {Oliver}, \citenamefont
  {Pettifor}, \citenamefont {Porfyrakis}, \citenamefont {Reina}, \citenamefont
  {Rice}, \citenamefont {Smith}, \citenamefont {Taylor}, \citenamefont
  {Williams}, \citenamefont {Adelmann}, \citenamefont {Mariette},\ and\
  \citenamefont {Hamers}}]{SSQC4}%
  \BibitemOpen
  \bibfield  {author} {\bibinfo {author} {\bibfnamefont {A.}~\bibnamefont
  {Ardavan}}, \bibinfo {author} {\bibfnamefont {M.}~\bibnamefont {Austwick}},
  \bibinfo {author} {\bibfnamefont {S.~C.}\ \bibnamefont {Benjamin}}, \bibinfo
  {author} {\bibfnamefont {G.~A.~D.}\ \bibnamefont {Briggs}}, \bibinfo {author}
  {\bibfnamefont {T.~J.~S.}\ \bibnamefont {Denni~s}}, \bibinfo {author}
  {\bibfnamefont {A.}~\bibnamefont {Ferguson}}, \bibinfo {author}
  {\bibfnamefont {D.~G.}\ \bibnamefont {Hasko}}, \bibinfo {author}
  {\bibfnamefont {M.}~\bibnamefont {Kanai}}, \bibinfo {author} {\bibfnamefont
  {A.~N.}\ \bibnamefont {Khlobystov}}, \bibinfo {author} {\bibfnamefont
  {B.~W.}\ \bibnamefont {Lovett}}, \bibinfo {author} {\bibfnamefont {G.~W.}\
  \bibnamefont {Morley}}, \bibinfo {author} {\bibfnamefont {R.~A.}\
  \bibnamefont {Oliver}}, \bibinfo {author} {\bibfnamefont {D.~G.}\
  \bibnamefont {Pettifor}}, \bibinfo {author} {\bibfnamefont {K.}~\bibnamefont
  {Porfyrakis}}, \bibinfo {author} {\bibfnamefont {J.~H.}\ \bibnamefont
  {Reina}}, \bibinfo {author} {\bibfnamefont {J.~H.}\ \bibnamefont {Rice}},
  \bibinfo {author} {\bibfnamefont {J.~D.}\ \bibnamefont {Smith}}, \bibinfo
  {author} {\bibfnamefont {R.~A.}\ \bibnamefont {Taylor}}, \bibinfo {author}
  {\bibfnamefont {D.~A.}\ \bibnamefont {Williams}}, \bibinfo {author}
  {\bibfnamefont {C.}~\bibnamefont {Adelmann}}, \bibinfo {author}
  {\bibfnamefont {H.}~\bibnamefont {Mariette}}, \ and\ \bibinfo {author}
  {\bibfnamefont {R.~J.}\ \bibnamefont {Hamers}},\ }\href {\doibase
  10.1098/rsta.2003.1214} {\bibfield  {journal} {\bibinfo  {journal} {Phil.
  Trans. R. Soc. A Math Phys. Eng. Sci.}\ }\textbf {\bibinfo {volume} {361}},\
  \bibinfo {pages} {1473} (\bibinfo {year} {2003})}\BibitemShut {NoStop}%
\bibitem [{\citenamefont {Feshbach}(1958)}]{Feshbach}%
  \BibitemOpen
  \bibfield  {author} {\bibinfo {author} {\bibfnamefont {H.}~\bibnamefont
  {Feshbach}},\ }\href@noop {} {\bibfield  {journal} {\bibinfo  {journal} {Ann.
  Phys. (N.~Y.)}\ }\textbf {\bibinfo {volume} {5}},\ \bibinfo {pages} {357}
  (\bibinfo {year} {1958})}\BibitemShut {NoStop}%
\bibitem [{\citenamefont {Sokolov}\ and\ \citenamefont
  {Zelevinsky}(1989)}]{SOKOLOV1}%
  \BibitemOpen
  \bibfield  {author} {\bibinfo {author} {\bibfnamefont {V.}~\bibnamefont
  {Sokolov}}\ and\ \bibinfo {author} {\bibfnamefont {V.}~\bibnamefont
  {Zelevinsky}},\ }\href {\doibase
  http://dx.doi.org/10.1016/0375-9474(89)90558-7} {\bibfield  {journal}
  {\bibinfo  {journal} {Nucl. Phys. A}\ }\textbf {\bibinfo {volume} {504}},\
  \bibinfo {pages} {562 } (\bibinfo {year} {1989})}\BibitemShut {NoStop}%
\bibitem [{\citenamefont {Sokolov}\ and\ \citenamefont
  {Zelevinsky}(1992)}]{SOKOLOV2}%
  \BibitemOpen
  \bibfield  {author} {\bibinfo {author} {\bibfnamefont {V.}~\bibnamefont
  {Sokolov}}\ and\ \bibinfo {author} {\bibfnamefont {V.}~\bibnamefont
  {Zelevinsky}},\ }\href {\doibase
  http://dx.doi.org/10.1016/0003-4916(92)90180-T} {\bibfield  {journal}
  {\bibinfo  {journal} {Ann. Phys.}\ }\textbf {\bibinfo {volume} {216}},\
  \bibinfo {pages} {323 } (\bibinfo {year} {1992})}\BibitemShut {NoStop}%
\bibitem [{\citenamefont {Tayebi}\ and\ \citenamefont
  {Zelevinsky}(2014)}]{Tayebi1}%
  \BibitemOpen
  \bibfield  {author} {\bibinfo {author} {\bibfnamefont {A.}~\bibnamefont
  {Tayebi}}\ and\ \bibinfo {author} {\bibfnamefont {V.}~\bibnamefont
  {Zelevinsky}},\ }\href@noop {} {\bibfield  {journal} {\bibinfo  {journal}
  {AIP Conf. Proc.}\ }\textbf {\bibinfo {volume} {1619}} (\bibinfo {year}
  {2014})}\BibitemShut {NoStop}%
\bibitem [{\citenamefont {Ziletti}\ \emph {et~al.}(2012)\citenamefont
  {Ziletti}, \citenamefont {Borgonovi}, \citenamefont {Celardo}, \citenamefont
  {Izrailev}, \citenamefont {Kaplan},\ and\ \citenamefont
  {Zelevinsky}}]{Ziletti}%
  \BibitemOpen
  \bibfield  {author} {\bibinfo {author} {\bibfnamefont {A.}~\bibnamefont
  {Ziletti}}, \bibinfo {author} {\bibfnamefont {F.}~\bibnamefont {Borgonovi}},
  \bibinfo {author} {\bibfnamefont {G.~L.}\ \bibnamefont {Celardo}}, \bibinfo
  {author} {\bibfnamefont {F.~M.}\ \bibnamefont {Izrailev}}, \bibinfo {author}
  {\bibfnamefont {L.}~\bibnamefont {Kaplan}}, \ and\ \bibinfo {author}
  {\bibfnamefont {V.~G.}\ \bibnamefont {Zelevinsky}},\ }\href {\doibase
  10.1103/PhysRevB.85.052201} {\bibfield  {journal} {\bibinfo  {journal} {Phys.
  Rev. B}\ }\textbf {\bibinfo {volume} {85}},\ \bibinfo {pages} {052201}
  (\bibinfo {year} {2012})}\BibitemShut {NoStop}%
\bibitem [{\citenamefont {Greenberg}\ and\ \citenamefont
  {Shtygashev}(2015)}]{Greenberg}%
  \BibitemOpen
  \bibfield  {author} {\bibinfo {author} {\bibfnamefont {Y.~S.}\ \bibnamefont
  {Greenberg}}\ and\ \bibinfo {author} {\bibfnamefont {A.~A.}\ \bibnamefont
  {Shtygashev}},\ }\href {\doibase 10.1103/PhysRevA.92.063835} {\bibfield
  {journal} {\bibinfo  {journal} {Phys. Rev. A}\ }\textbf {\bibinfo {volume}
  {92}},\ \bibinfo {pages} {063835} (\bibinfo {year} {2015})}\BibitemShut
  {NoStop}%
\bibitem [{\citenamefont {Auerbach}\ and\ \citenamefont
  {Zelevinsky}(2011)}]{Auerbach}%
  \BibitemOpen
  \bibfield  {author} {\bibinfo {author} {\bibfnamefont {N.}~\bibnamefont
  {Auerbach}}\ and\ \bibinfo {author} {\bibfnamefont {V.}~\bibnamefont
  {Zelevinsky}},\ }\href {http://stacks.iop.org/0034-4885/74/i=10/a=106301}
  {\bibfield  {journal} {\bibinfo  {journal} {Rep. Prog. Phys.}\ }\textbf
  {\bibinfo {volume} {74}},\ \bibinfo {pages} {106301} (\bibinfo {year}
  {2011})}\BibitemShut {NoStop}%
\bibitem [{\citenamefont {Zhang}\ \emph {et~al.}(2012)\citenamefont {Zhang},
  \citenamefont {Ye}, \citenamefont {Wang}, \citenamefont {Park}, \citenamefont
  {Bartal}, \citenamefont {Mrejen}, \citenamefont {Yin},\ and\ \citenamefont
  {Zhang}}]{PlasmonicArray}%
  \BibitemOpen
  \bibfield  {author} {\bibinfo {author} {\bibfnamefont {S.}~\bibnamefont
  {Zhang}}, \bibinfo {author} {\bibfnamefont {Z.}~\bibnamefont {Ye}}, \bibinfo
  {author} {\bibfnamefont {Y.}~\bibnamefont {Wang}}, \bibinfo {author}
  {\bibfnamefont {Y.}~\bibnamefont {Park}}, \bibinfo {author} {\bibfnamefont
  {G.}~\bibnamefont {Bartal}}, \bibinfo {author} {\bibfnamefont
  {M.}~\bibnamefont {Mrejen}}, \bibinfo {author} {\bibfnamefont
  {X.}~\bibnamefont {Yin}}, \ and\ \bibinfo {author} {\bibfnamefont
  {X.}~\bibnamefont {Zhang}},\ }\href {\doibase 10.1103/PhysRevLett.109.193902}
  {\bibfield  {journal} {\bibinfo  {journal} {Phys. Rev. Lett.}\ }\textbf
  {\bibinfo {volume} {109}},\ \bibinfo {pages} {193902} (\bibinfo {year}
  {2012})}\BibitemShut {NoStop}%
\bibitem [{\citenamefont {Celardo}\ \emph {et~al.}(2012)\citenamefont
  {Celardo}, \citenamefont {Borgonovi}, \citenamefont {Merkli}, \citenamefont
  {Tsifrinovich},\ and\ \citenamefont {Berman}}]{CelardoBiology}%
  \BibitemOpen
  \bibfield  {author} {\bibinfo {author} {\bibfnamefont {G.~L.}\ \bibnamefont
  {Celardo}}, \bibinfo {author} {\bibfnamefont {F.}~\bibnamefont {Borgonovi}},
  \bibinfo {author} {\bibfnamefont {M.}~\bibnamefont {Merkli}}, \bibinfo
  {author} {\bibfnamefont {V.~I.}\ \bibnamefont {Tsifrinovich}}, \ and\
  \bibinfo {author} {\bibfnamefont {G.~P.}\ \bibnamefont {Berman}},\ }\href
  {\doibase 10.1021/jp302627w} {\bibfield  {journal} {\bibinfo  {journal} {J.
  Phys. Chem. C}\ }\textbf {\bibinfo {volume} {116}},\ \bibinfo {pages} {22105}
  (\bibinfo {year} {2012})}\BibitemShut {NoStop}%
\bibitem [{\citenamefont {Greenberg}\ \emph {et~al.}(2013)\citenamefont
  {Greenberg}, \citenamefont {Merrigan}, \citenamefont {Tayebi},\ and\
  \citenamefont {Zelevinsky}}]{Tayebi2}%
  \BibitemOpen
  \bibfield  {author} {\bibinfo {author} {\bibfnamefont {Y.~S.}\ \bibnamefont
  {Greenberg}}, \bibinfo {author} {\bibfnamefont {C.}~\bibnamefont {Merrigan}},
  \bibinfo {author} {\bibfnamefont {A.}~\bibnamefont {Tayebi}}, \ and\ \bibinfo
  {author} {\bibfnamefont {V.}~\bibnamefont {Zelevinsky}},\ }\href {\doibase
  10.1140/epjb/e2013-40190-4} {\bibfield  {journal} {\bibinfo  {journal} {Eur.
  Phys. J. B}\ }\textbf {\bibinfo {volume} {86}},\ \bibinfo {pages} {1}
  (\bibinfo {year} {2013})}\BibitemShut {NoStop}%
\bibitem [{\citenamefont {Kwapi\ifmmode~\acute{n}\else \'{n}\fi{}ski}\ and\
  \citenamefont {Taranko}(2012)}]{Kwapi}%
  \BibitemOpen
  \bibfield  {author} {\bibinfo {author} {\bibfnamefont {T.}~\bibnamefont
  {Kwapi\ifmmode~\acute{n}\else \'{n}\fi{}ski}}\ and\ \bibinfo {author}
  {\bibfnamefont {R.}~\bibnamefont {Taranko}},\ }\href {\doibase
  10.1103/PhysRevA.86.052338} {\bibfield  {journal} {\bibinfo  {journal} {Phys.
  Rev. A}\ }\textbf {\bibinfo {volume} {86}},\ \bibinfo {pages} {052338}
  (\bibinfo {year} {2012})}\BibitemShut {NoStop}%
\bibitem [{\citenamefont {Baines}\ \emph {et~al.}(2012)\citenamefont {Baines},
  \citenamefont {Meunier}, \citenamefont {Mailly}, \citenamefont {Wieck},
  \citenamefont {B\"auerle}, \citenamefont {Saminadayar}, \citenamefont
  {Cornaglia}, \citenamefont {Usaj}, \citenamefont {Balseiro},\ and\
  \citenamefont {Feinberg}}]{expermintnu}%
  \BibitemOpen
  \bibfield  {author} {\bibinfo {author} {\bibfnamefont {D.~Y.}\ \bibnamefont
  {Baines}}, \bibinfo {author} {\bibfnamefont {T.}~\bibnamefont {Meunier}},
  \bibinfo {author} {\bibfnamefont {D.}~\bibnamefont {Mailly}}, \bibinfo
  {author} {\bibfnamefont {A.~D.}\ \bibnamefont {Wieck}}, \bibinfo {author}
  {\bibfnamefont {C.}~\bibnamefont {B\"auerle}}, \bibinfo {author}
  {\bibfnamefont {L.}~\bibnamefont {Saminadayar}}, \bibinfo {author}
  {\bibfnamefont {P.~S.}\ \bibnamefont {Cornaglia}}, \bibinfo {author}
  {\bibfnamefont {G.}~\bibnamefont {Usaj}}, \bibinfo {author} {\bibfnamefont
  {C.~A.}\ \bibnamefont {Balseiro}}, \ and\ \bibinfo {author} {\bibfnamefont
  {D.}~\bibnamefont {Feinberg}},\ }\href {\doibase 10.1103/PhysRevB.85.195117}
  {\bibfield  {journal} {\bibinfo  {journal} {Phys. Rev. B}\ }\textbf {\bibinfo
  {volume} {85}},\ \bibinfo {pages} {195117} (\bibinfo {year}
  {2012})}\BibitemShut {NoStop}%
\bibitem [{\citenamefont {Fringes}\ \emph {et~al.}(2012)\citenamefont
  {Fringes}, \citenamefont {Volk}, \citenamefont {Terrés}, \citenamefont
  {Dauber}, \citenamefont {Engels}, \citenamefont {Trellenkamp},\ and\
  \citenamefont {Stampfer}}]{expermintnu2}%
  \BibitemOpen
  \bibfield  {author} {\bibinfo {author} {\bibfnamefont {S.}~\bibnamefont
  {Fringes}}, \bibinfo {author} {\bibfnamefont {C.}~\bibnamefont {Volk}},
  \bibinfo {author} {\bibfnamefont {B.}~\bibnamefont {Terrés}}, \bibinfo
  {author} {\bibfnamefont {J.}~\bibnamefont {Dauber}}, \bibinfo {author}
  {\bibfnamefont {S.}~\bibnamefont {Engels}}, \bibinfo {author} {\bibfnamefont
  {S.}~\bibnamefont {Trellenkamp}}, \ and\ \bibinfo {author} {\bibfnamefont
  {C.}~\bibnamefont {Stampfer}},\ }\href {\doibase 10.1002/pssc.201100340}
  {\bibfield  {journal} {\bibinfo  {journal} {Phys. Status Solidi C}\ }\textbf
  {\bibinfo {volume} {9}},\ \bibinfo {pages} {169} (\bibinfo {year}
  {2012})}\BibitemShut {NoStop}%
\bibitem [{\citenamefont {Simmons}\ \emph {et~al.}(2009)\citenamefont
  {Simmons}, \citenamefont {Thalakulam}, \citenamefont {Rosemeyer},
  \citenamefont {Bael}, \citenamefont {Sackmann}, \citenamefont {Savage},
  \citenamefont {Lagally}, \citenamefont {Joynt}, \citenamefont {Friesen},
  \citenamefont {Coppersmith},\ and\ \citenamefont {Eriksson}}]{Simmons}%
  \BibitemOpen
  \bibfield  {author} {\bibinfo {author} {\bibfnamefont {C.~B.}\ \bibnamefont
  {Simmons}}, \bibinfo {author} {\bibfnamefont {M.}~\bibnamefont {Thalakulam}},
  \bibinfo {author} {\bibfnamefont {B.~M.}\ \bibnamefont {Rosemeyer}}, \bibinfo
  {author} {\bibfnamefont {B.~J.~V.}\ \bibnamefont {Bael}}, \bibinfo {author}
  {\bibfnamefont {E.~K.}\ \bibnamefont {Sackmann}}, \bibinfo {author}
  {\bibfnamefont {D.~E.}\ \bibnamefont {Savage}}, \bibinfo {author}
  {\bibfnamefont {M.~G.}\ \bibnamefont {Lagally}}, \bibinfo {author}
  {\bibfnamefont {R.}~\bibnamefont {Joynt}}, \bibinfo {author} {\bibfnamefont
  {M.}~\bibnamefont {Friesen}}, \bibinfo {author} {\bibfnamefont {S.~N.}\
  \bibnamefont {Coppersmith}}, \ and\ \bibinfo {author} {\bibfnamefont {M.~A.}\
  \bibnamefont {Eriksson}},\ }\href {\doibase 10.1021/nl9014974} {\bibfield
  {journal} {\bibinfo  {journal} {Nano Letters}\ }\textbf {\bibinfo {volume}
  {9}},\ \bibinfo {pages} {3234} (\bibinfo {year} {2009})}\BibitemShut
  {NoStop}%
\bibitem [{\citenamefont {DiCarlo}\ \emph {et~al.}(2004)\citenamefont
  {DiCarlo}, \citenamefont {Lynch}, \citenamefont {Johnson}, \citenamefont
  {Childress}, \citenamefont {Crockett}, \citenamefont {Marcus}, \citenamefont
  {Hanson},\ and\ \citenamefont {Gossard}}]{DiCarlo}%
  \BibitemOpen
  \bibfield  {author} {\bibinfo {author} {\bibfnamefont {L.}~\bibnamefont
  {DiCarlo}}, \bibinfo {author} {\bibfnamefont {H.~J.}\ \bibnamefont {Lynch}},
  \bibinfo {author} {\bibfnamefont {A.~C.}\ \bibnamefont {Johnson}}, \bibinfo
  {author} {\bibfnamefont {L.~I.}\ \bibnamefont {Childress}}, \bibinfo {author}
  {\bibfnamefont {K.}~\bibnamefont {Crockett}}, \bibinfo {author}
  {\bibfnamefont {C.~M.}\ \bibnamefont {Marcus}}, \bibinfo {author}
  {\bibfnamefont {M.~P.}\ \bibnamefont {Hanson}}, \ and\ \bibinfo {author}
  {\bibfnamefont {A.~C.}\ \bibnamefont {Gossard}},\ }\href {\doibase
  10.1103/PhysRevLett.92.226801} {\bibfield  {journal} {\bibinfo  {journal}
  {Phys. Rev. Lett.}\ }\textbf {\bibinfo {volume} {92}},\ \bibinfo {pages}
  {226801} (\bibinfo {year} {2004})}\BibitemShut {NoStop}%
\bibitem [{\citenamefont {Celardo}\ and\ \citenamefont
  {Kaplan}(2009)}]{Celardo}%
  \BibitemOpen
  \bibfield  {author} {\bibinfo {author} {\bibfnamefont {G.~L.}\ \bibnamefont
  {Celardo}}\ and\ \bibinfo {author} {\bibfnamefont {L.}~\bibnamefont
  {Kaplan}},\ }\href {\doibase 10.1103/PhysRevB.79.155108} {\bibfield
  {journal} {\bibinfo  {journal} {Phys. Rev. B}\ }\textbf {\bibinfo {volume}
  {79}},\ \bibinfo {pages} {155108} (\bibinfo {year} {2009})}\BibitemShut
  {NoStop}%
\bibitem [{\citenamefont {Dicke}(1954)}]{DickeQO}%
  \BibitemOpen
  \bibfield  {author} {\bibinfo {author} {\bibfnamefont {R.~H.}\ \bibnamefont
  {Dicke}},\ }\href {\doibase 10.1103/PhysRev.93.99} {\bibfield  {journal}
  {\bibinfo  {journal} {Phys. Rev.}\ }\textbf {\bibinfo {volume} {93}},\
  \bibinfo {pages} {99} (\bibinfo {year} {1954})}\BibitemShut {NoStop}%
\bibitem [{\citenamefont {von Brentano}(1996)}]{vonBrentano}%
  \BibitemOpen
  \bibfield  {author} {\bibinfo {author} {\bibfnamefont {P.}~\bibnamefont {von
  Brentano}},\ }\href {\doibase http://dx.doi.org/10.1016/0370-1573(95)00027-5}
  {\bibfield  {journal} {\bibinfo  {journal} {Phys. Rep.}\ }\textbf {\bibinfo
  {volume} {264}},\ \bibinfo {pages} {57} (\bibinfo {year} {1996})}\BibitemShut
  {NoStop}%
\bibitem [{\citenamefont {Volya}\ and\ \citenamefont
  {Zelevinsky}(2005)}]{volyazel}%
  \BibitemOpen
  \bibfield  {author} {\bibinfo {author} {\bibfnamefont {A.}~\bibnamefont
  {Volya}}\ and\ \bibinfo {author} {\bibfnamefont {V.}~\bibnamefont
  {Zelevinsky}},\ }\href {\doibase http://dx.doi.org/10.1063/1.1996889}
  {\bibfield  {journal} {\bibinfo  {journal} {AIP Conf. Proc.}\ }\textbf
  {\bibinfo {volume} {777}},\ \bibinfo {pages} {229} (\bibinfo {year}
  {2005})}\BibitemShut {NoStop}%
\bibitem [{\citenamefont {Haken}\ and\ \citenamefont
  {Strobl}(1973)}]{HakenStrobl}%
  \BibitemOpen
  \bibfield  {author} {\bibinfo {author} {\bibfnamefont {H.}~\bibnamefont
  {Haken}}\ and\ \bibinfo {author} {\bibfnamefont {G.}~\bibnamefont {Strobl}},\
  }\href {\doibase 10.1007/BF01399723} {\bibfield  {journal} {\bibinfo
  {journal} {Z. Phys.}\ }\textbf {\bibinfo {volume} {262}},\ \bibinfo {pages}
  {135} (\bibinfo {year} {1973})}\BibitemShut {NoStop}%
\bibitem [{\citenamefont {Grad}\ \emph {et~al.}(1988)\citenamefont {Grad},
  \citenamefont {Hernandez},\ and\ \citenamefont {Mukamel}}]{Grad}%
  \BibitemOpen
  \bibfield  {author} {\bibinfo {author} {\bibfnamefont {J.}~\bibnamefont
  {Grad}}, \bibinfo {author} {\bibfnamefont {G.}~\bibnamefont {Hernandez}}, \
  and\ \bibinfo {author} {\bibfnamefont {S.}~\bibnamefont {Mukamel}},\ }\href
  {\doibase 10.1103/PhysRevA.37.3835} {\bibfield  {journal} {\bibinfo
  {journal} {Phys. Rev. A}\ }\textbf {\bibinfo {volume} {37}},\ \bibinfo
  {pages} {3835} (\bibinfo {year} {1988})}\BibitemShut {NoStop}%
\bibitem [{\citenamefont {Cheng}\ and\ \citenamefont {Silbey}(2004)}]{Cheng}%
  \BibitemOpen
  \bibfield  {author} {\bibinfo {author} {\bibfnamefont {Y.~C.}\ \bibnamefont
  {Cheng}}\ and\ \bibinfo {author} {\bibfnamefont {R.~J.}\ \bibnamefont
  {Silbey}},\ }\href {\doibase 10.1103/PhysRevA.69.052325} {\bibfield
  {journal} {\bibinfo  {journal} {Phys. Rev. A}\ }\textbf {\bibinfo {volume}
  {69}},\ \bibinfo {pages} {052325} (\bibinfo {year} {2004})}\BibitemShut
  {NoStop}%
\bibitem [{\citenamefont {Celardo}\ \emph {et~al.}(2014)\citenamefont
  {Celardo}, \citenamefont {Poli}, \citenamefont {Lussardi},\ and\
  \citenamefont {Borgonovi}}]{CelardoR}%
  \BibitemOpen
  \bibfield  {author} {\bibinfo {author} {\bibfnamefont {G.~L.}\ \bibnamefont
  {Celardo}}, \bibinfo {author} {\bibfnamefont {P.}~\bibnamefont {Poli}},
  \bibinfo {author} {\bibfnamefont {L.}~\bibnamefont {Lussardi}}, \ and\
  \bibinfo {author} {\bibfnamefont {F.}~\bibnamefont {Borgonovi}},\ }\href
  {\doibase 10.1103/PhysRevB.90.085142} {\bibfield  {journal} {\bibinfo
  {journal} {Phys. Rev. B}\ }\textbf {\bibinfo {volume} {90}},\ \bibinfo
  {pages} {085142} (\bibinfo {year} {2014})}\BibitemShut {NoStop}%
\bibitem [{\citenamefont {Reineker}(1982)}]{Reineker}%
  \BibitemOpen
  \bibfield  {author} {\bibinfo {author} {\bibfnamefont {P.}~\bibnamefont
  {Reineker}},\ }\href@noop {} {\emph {\bibinfo {title} {Exciton dynamics in
  molecular crystals and aggregates}}}\ (\bibinfo  {publisher} {Springer,
  Berlin-Heidelberg},\ \bibinfo {year} {1982})\BibitemShut {NoStop}%
\end{thebibliography}%

\end{document}